\documentclass[twocolumn,preprintnumbers,amsmath,amssymb]{revtex4}
\usepackage{graphicx}
\usepackage{dcolumn}
\usepackage{bm}
\usepackage{color}

\def\ro{\rm o}

\begin{document}


\title[Optical tweezers absolute calibration]{Optical tweezers absolute calibration}

\author{R S Dutra$^{1,2,3}$, N B Viana$^{2,3}$, P A Maia Neto$^{2,3}$ and H M Nussenzveig$^{2,3}$}
\address{$^1$Instituto Federal de Educa\c{c}\~ao, Ci\^encia e Tecnologia, Rua Sebasti\~ao de Lacerda, Paracambi, RJ, 26600-000, Brasil}
\address{$^2$LPO-COPEA, Instituto de Ci\^encias Biom\'edicas, Universidade Federal do Rio de Janeiro, Rio de Janeiro, RJ, 21941-590, Brasil}
\address{$^3$Instituto de F\'isica, Universidade Federal do Rio de Janeiro, Caixa Postal 68528, Rio de Janeiro, RJ, 21941-972, Brasil
}

\date{\today}

\begin{abstract}

 Optical tweezers are highly versatile laser traps for neutral microparticles, with fundamental applications in physics and in single molecule cell biology. Force measurements are performed by converting the stiffness response to displacement of trapped transparent microspheres, employed as force transducers. Usually, calibration is indirect, by comparison with fluid drag forces. This can lead to discrepancies by sizable factors. Progress achieved in a program aiming at absolute calibration, conducted over the past fifteen years, is briefly reviewed.	Here we overcome its last major obstacle, a theoretical overestimation of the peak stiffness, within the most employed range for applications, and we perform experimental validation. The discrepancy is traced to the effect of primary aberrations of the optical system, which are now included in the theory. All required experimental parameters are readily accessible. Astigmatism, the dominant effect, is measured by analyzing reflected images of the focused laser spot, adapting frequently employed video microscopy techniques. Combined with interface spherical aberration, it reveals a previously unknown window of instability for trapping. Comparison with experimental data leads to an overall agreement within error bars, with no fitting, for a broad range of microsphere radii, from the Rayleigh regime to the ray optics one, for different polarizations and trapping heights, including all commonly employed parameter domains. Besides signalling full first-principles theoretical understanding of optical tweezers operation, the results may lead to improved instrument design and control over experiments, as well as to an extended domain of applicability, allowing reliable force measurements, in principle, from femtonewtons to nanonewtons.
   
\end{abstract}


\maketitle

\section{Introduction}

Optical tweezers (OT), invented in 1986 \cite{Ashkin86}, are laser traps for neutral microscopic particles, with a vast range of applications in physics and biology: a 2006 review~\cite{Phys} lists $\sim 10^3$ publications. Recent applications to fundamental physics include an experimental realization of Szilard's demon
 \cite{Toyave2010} and the first experimental proof of Landauer's principle \cite{Berut2012}. In cell biology, OT have paved the way to pioneering quantitative measurements of basic interactions in living cells, ``one molecule at a time'' \cite{Bustamante2011,Block2011}.

For biological applications, one employs near-infrared laser light, within a transparency window for the water contained in cells, to avoid heat damage. A transparent microsphere is employed as a handle and force transducer. The microsphere is pulled toward the diffraction-limited laser focus by the gradient force, which must overcome the opposing radiation pressure, thus requiring a strongly focused beam. The beam is focused through the microscope objective. To maintain the live biological sample, it is usually immersed in water solutions, within a chamber with controlled temperature and carbon dioxide pressure. 
For
a schematic diagram of a typical set-up see \cite{Neuman2004}.

The object of interest is attached to the trapped microsphere, through which the force is applied, usually transverse to the beam. For sufficiently small microsphere displacements from equilibrium, the response is linear both in displacement and in laser power, so that it suffices to calibrate the transverse trap stiffness per unit power, measuring the displacement to determine the force.

Stiffness calibration is usually based on comparisons with fluid drag forces \cite{Simmons96} or on detection of thermal fluctuations \cite{Berg-Sorensen2004}
 by assuming a  known drag coefficient. 
Alternatively, measuring the power spectra under a sinusoidal motion of the microscope translational stage allows for an independent calibration of the drag force on the trapped particle~\cite{Tolic-Norrelykke2006}.
 In cell biology, forces may need to be measured under complicated boundary conditions, at micrometer distances from the bottom of the sample chamber. Results at different laboratories can disagree even by an order of magnitude (e. g., \cite{LPO}).

In the present work, we demonstrate an absolute calibration of stiffness, based on a 
 careful control of all relevant trap parameters  \cite{NathanPRE} and on an accurate realistic theory of the trapping force,  yielding the stiffness in terms of experimentally accessible data. The basic ingredients of such a theory are the description of the strongly focused laser beam and of its interaction with the microsphere.

Early treatments \cite{Ashkin86} of the interaction were confined to the Rayleigh regime (microsphere radius $a$ below $0.1\, \mu{\rm m}$), in which the stiffness
 grows like $a^3$, and to the geometrical optics limit \cite{Roosen1979,Ashkin1992}
$a\gg \lambda,$
 where $\lambda\sim 1\, \mu{\rm m}$ is the laser wavelength. In usual experiments, 
 $a$ 
  is  $\stackrel{<}{\scriptscriptstyle\sim}  1\, \mu{\rm m},$ in the Mie regime.
   In the widely referenced ``generalized Lorenz-Mie theory'' \cite{Gouesbet85,Barton89}, however, the trapping beam was described in terms of perturbative corrections to a paraxial Gaussian model, which has been shown \cite{Ganic2004} to be an incorrect representation of a strongly focused beam. A proper representation of such a beam \cite{Ashkin1992} is the electromagnetic generalization \cite{RichardsWolf59} of Debye's classic scalar model \cite{Debye}. 
   A more detailed overview of other proposals is given in \cite{NathanPRE}.
   
   The generalized Debye representation, combined with Mie theory, and taking due account of the Abbe sine condition, was first applied to 
   the axial stiffness \cite{MaiaNeto00}. 
   This MD (Mie-Debye) theory predicts rapid oscillations in $a /\lambda,$  arising from interference between contributions from the sphere edges for spectral angular components. Related oscillations have been detected in optical trapping of water droplets \cite{Guillon}. As is expected in semiclassical scattering \cite{Berry}, averaging over oscillations,	for	$a \gg \lambda,$  yields	the	geometrical	optics	result,	which	decays asymptotically like $1/a,$ 
   as follows from dimensional analysis. Previous theories did not show oscillations and had incorrect asymptotic behavior.
   
The extension of MD theory to the more relevant transverse stiffness \cite{Mazolli03}, with similar features, differed from available experimental data by an apparent overall displacement. This was traced back to its disregard of interface spherical aberration,
 the defocussing of the laser beam by refraction at the  interface
 between the glass slide and the water in the sample chamber~\cite{Torok95}. 
 Inclusion of this effect led to the MDSA (Mie-Debye-Interface Spherical Aberration) theory 
   \cite{NathanPRE}.

Extensive experimental tests of the MDSA theory for different OT setups
\cite{NathanPRE,NathanAPL,Dutra}
 showed good agreement with its predictions in the range $a > \lambda,$ for the trapping threshold, location of the stiffness peak, stiffness degradation with height in the sample chamber, and ``hopping''  between multiple equilibrium points. 
 Recent extensions include modeling counterpropagating dual-beam \cite{vanderHorst} and aerosol optical traps \cite{Burnham2011}.
However, under the usual conditions of an overfilled high numerical aperture (NA) objective,
 MDSA leads to a huge overestimation of the stiffness in the interval $a\stackrel{<}{\scriptscriptstyle\sim}  \lambda/2,$
 where the predicted stiffness is maximal.
  This is precisely the size domain of greatest importance for practical applications, thus compromising the validity of  MDSA for absolute calibration.
  It	was	conjectured in \cite{NathanPRE} that additional optical aberrations of the microscope objective 
  could be responsible for the stiffness reduction, by degrading the focus.

In the present work, we investigate in detail the effects of all primary aberrations on the optical trapping force. 
Building on our previous theoretical work, 
we develop a new model, denoted as MDSA+, that takes into account the presence of primary aberrations of the focused trapping beam in addition to the interface spherical aberration.

We show that one additional optical aberration, astigmatism, is the main effect responsible for the transverse stiffness degradation in the range 
$a\stackrel{<}{\scriptscriptstyle\sim}  \lambda/2.$ We independently characterize the astigmatism of our OT setup and plug the results into MDSA+ theory. We find agreement with the experimental data within error bars, with no fitting procedure. The success of such blind comparison is of particular importance, given that astigmatism is always present to some degree in typical OT setups (see for instance \cite{Roichman06}). It also demonstrates that absolute calibration of the trap stiffness can be achieved, provided that all relevant experimental parameters, including the astigmatism, are carefully characterized.

Preliminary results for the case of circular polarization were briefly reported in Ref. \cite{Dutra2012}.
 Here we present a comprehensive account of the effects of all primary aberrations on the optical force field. We choose to present 
 the most common case of linear polarization so as to provide more useful guidelines for typical optical tweezers setups.

The paper is organized as follows: in Sec. II we develop MDSA+ and consider  numerical examples, taking each primary aberration separately 
(explicit formulas are given in Appendix A). 
Sec. III is dedicated to the characterization of astigmatism in our typical OT setup. We compare experimental data with theoretical predictions for the trap stiffness in Sec. IV. Concluding remarks are presented in Sec. V. The main conclusion is that absolute calibration of optical tweezers has finally been achieved and that it should lead to significant practical consequences. Appendix B provides a short guide to the implementation of absolute calibration.

 \section{MDSA+ theory of the optical force in the presence of aberrations}
 
 \subsection{General formalism\label{general}}
 
 In this section, we derive formal results for the optical force in the presence of aberrations. In  the typical optical tweezer setup,
  the trapping laser beam is focused by a high numerical-aperture (NA)  oil-immersion objective into a sample region filled with water. 
 The effect of the spherical aberration introduced by the glass-water planar interface was already analyzed in detail in Ref.~\cite{Dutra}.
   Here we also take into account additional optical aberrations introduced by the objective itself and by  optical
 elements along the optical path before the objective. 
 
 We assume that the trapping laser beam at the entrance port (aperture radius $R_0$) of the infinity-corrected microscope objective  (focal length $f$) has 
 amplitude $E_p$,
 waist 
 $w_0$  and
is linearly polarized along the $x$-axis. 
We employ the Seidel formalism for the aberrations  \cite{ Born&Wolf}. 
Among the  Seidel primary aberrations, we expect field curvature and distortion to keep 
Êthe	
 Êthree-dimensional intensity distribution around the focal region
approximately unchanged, except for a 
 global spatial translation (displacement theorem \cite{ Born&Wolf}). Thus, we focus on the effects of spherical aberration, coma and astigmatism. 
 
To include these three  primary  aberrations  into our theoretical model, we introduce the corresponding phase for each plane wave component
associated to a given angle 
 $(\theta,\varphi)$
(in spherical coordinates) into the Debye-type angular spectrum 
 representation of the focused beam. 
We assume that the objective  satisfies the usual sine condition and we write the focused electric field as (origin at the paraxial focus)
\begin{eqnarray} 
\mathbf{E}(\mathbf{r})&=&{E}_{0}\int_{0}^{2\pi}d\varphi\int_{0}^{\theta_{m}}d{\theta}\sin \theta \sqrt{\cos \theta}e^{-\gamma^2\sin^2\theta} 
\,T(\theta)\nonumber\\
\label{main1}
&&\times
 e^{i[\Phi_{\rm g-w}(\theta)
+\Phi_{\rm add}(\theta,\varphi)]}  e^{i\mathbf{k}_{\rm w}\cdot\mathbf{r}}\mathbf{\hat{x}^{\prime}}(\theta_{\rm w},\varphi_{\rm w}),\label{campofocalizado} 
\end{eqnarray}
with $\gamma = f/w_{0}$ and   ${E}_{0}=-(ik f/2\pi) E_{p} \exp\left(i k f\right).$
The wavevector 
 in the sample region has modulus $ k_{\rm w}=N k$ where  $N=n_{\rm w}/n$ is the relative refractive index for the glass-water interface
 and $k$ is the wavenumber in the glass medium of refractive index $n.$ The direction of  $\mathbf{k}_{\rm w}$ is defined by the spherical coordinates
  $(\theta_{\rm w},\varphi_{\rm w}),$
  where 
   $\theta_{\rm w}=\sin^{-1}(\sin\theta/N)$ (refraction angle) and $\varphi_{\rm w}=\varphi.$
 The Fresnel refraction amplitude
\begin{equation}
T(\theta)= \frac{2\cos\theta}{\cos\theta+N\cos\theta_{\rm w}}
\end{equation}
accounts for the amplitude transmission across the interface.
More importantly, Eq.~(\ref{main1}) contains the phase
\begin{equation} \label{aberration_interface}
\Phi_{\rm g-w}(\theta)=k\left( -\frac{L}{N}\cos\theta+NL\cos\theta_{\rm w}\right),
\end{equation}
proportional to the distance $L$ between the paraxial focus and the planar interface, accounting for the spherical aberration introduced by refraction at the glass-water interface.
The unit vector ${\mathbf{\hat x}}^{\prime}(\theta_{\rm w},\varphi_{\rm w})$
 in (\ref{main1}) is obtained from $\mathbf{\hat x}$ by rotation with Euler angles $(\varphi_{\rm w},\theta_{\rm w},-\varphi_{\rm w}):$
\[
{\mathbf{\hat x}}^{\prime}=\cos\varphi_{\rm w}\,\hat{\mbox{\boldmath$\theta$}}_{\rm w}-\sin\varphi_{\rm w} \hat{\mbox{\boldmath$\varphi$}}_{\rm w}.
\]

When the numerical aperture (NA) is larger than $n_{\rm w}$ (for instance for the popular $\mbox{NA}=1.4$ objectives), part of the angular spectrum 
exceeds the critical angle $\theta_{\rm cr}= \sin^{-1}(N)$ for total internal reflection, producing evanescent waves in the sample region. Here we assume that the 
trapped microsphere is several wavelengths away from the interface, allowing us to neglect the contribution of the evanescent sector. Thus, 
we limit the integration in  (\ref{main1}) to 
\[
 \theta_{m} =\mbox{min}\{\theta_{\rm cr}, \theta_0\equiv  \sin^{-1}(\mbox{NA}/n) \}.
\]

The main novelty in 
Eq. (\ref{main1}) is the  phase 
\begin{eqnarray}
\frac{\Phi_{\rm add}(\theta,\varphi)}{2\pi} &=&  A'_{\rm sa}\left( \frac{\sin\theta}{\sin\theta_{0}}\right)^4 +
  A'_{\rm c} \left(\frac{\sin \theta}{\sin \theta_{0}}\right)^3 \cos(\varphi - \phi_{\rm c})\nonumber
  \\
 &&  +\,  A'_{\rm a}\left( \frac{\sin \theta}{\sin \theta_{0}}\right)^2 
  \cos^2(\varphi - \phi_{\rm a}) , \label{funcaoaberracaoum} 
\end{eqnarray}
containing  the relevant  primary aberrations in the optical system (objective included).
In (\ref{funcaoaberracaoum}), 
$A'_{\rm sa},$ $ A'_{\rm c}$ and $A'_{\rm a}$ represent the amplitudes of  system spherical aberration (in addition to the one introduced by the glass-water interface), 
coma and astigmatism, respectively. 
The index sa is meant to distinguish optical system (objective and remaining optical elements, e.g., telescopic system) spherical aberration 
from interface 
 spherical aberration, already included in MDSA.	
The coma and astigmatism axes are defined by the angles  $\phi_{\rm c}$ and $\phi_{\rm a},$
measured with respect to the $x$ axis in the image space of the 
objective.

The scattered fields for each plane wave component in (\ref{main1}) are written in terms of Wigner rotation matrix elements
$d^j_{m,m'}(\theta_{\rm w})$ \cite{Edmonds}
 and 
Mie coefficients  $a_j$ and $b_j$ for electric and magnetic multipoles, respectively \cite{Bohren&Huffman}.
The integer variables $j\ge 1$ and  $m=-j,...,j$ represent the  total angular momentum $J^2$ (eigenvalues $j(j+1)$) and its axial component  $J_z,$
respectively.
After  
expanding the focused field (\ref{main1}) into multipoles, we evaluate the integral over the azimuth angle $\varphi$ 
and use 
 Graf's generalization of  Neumann's addition theorem for  cylindrical Bessel functions $J_n(x)$ \cite{Watson}. 
A partial-wave (multipole)
representation for the optical force is then derived from the Maxwell stress tensor \cite{Farsund}. 
Since the optical force $\mathbf{F}$  is proportional to the laser beam power  $P$ at the sample region, it is convenient to 
define the dimensionless vector efficiency factor \cite{Ashkin1992}
\begin{equation}
\mathbf{Q} = \frac{\mathbf{F}}{n_{\rm w}P/c}, \label{defQ}
\end{equation}
where  $c$
is the speed of light in vacuum. The cylindrical components of $\mathbf{Q}$ are given in  Appendix A, in terms 
of the incident field multipole coefficients
\begin{widetext}
\begin{eqnarray}\label{Gjm}
G^{(\sigma)}_{jm}(\rho,\phi,z)=\int_{0}^{\theta_m}d\theta\sin\theta\sqrt{\cos \theta}\,e^{-\gamma^2\sin^2\theta}\,
T(\theta)\,d_{m,\sigma}^{j}(\theta_{\rm w})\,g^{(\sigma)}_{m}(\rho,\phi,\theta)\exp\left\{i[\Phi_{\rm g-w}(\theta)+\Psi_{\rm add}(\theta)+k_{\rm w}\cos\theta_{\rm w} z ]\right\}
\end{eqnarray}
where 
$\sigma=\pm 1 $ 
denotes the photon helicity.
The phase 
\begin{equation}
\Psi_{\rm add}(\theta)= 2\pi A'_{\rm sa} \left(\frac{\sin \theta}{\sin \theta_{0}}\right)^4 + 
\pi A'_{\rm a} \left(\frac{\sin \theta}{\sin \theta_{0}}\right)^2
\end{equation}
accounts for additional spherical aberration and a residual field curvature arising from the Seidel astigmatism. 
The anisotropy introduced by astigmatism and coma
is contained in the function 
\begin{equation}
g^{(\sigma)}_{m}(\rho,\phi,\theta) = e^{i(m-\sigma)\alpha}\sum_{s=-\infty}^{\infty}(-i)^s 
J_{s}\biggl(\pi A'_{\rm a} \frac{\sin^2\theta}{\sin^2\theta_{0}}\biggl)\,
J_{2s+m-\sigma}(|{\cal Z}|)\,
e^{2is(\alpha+\phi_{\rm a}-\phi)}
 \label{function_f}
\end{equation}
\end{widetext}
where the coma parameters define the 
 complex quantity
\begin{equation}
{\cal Z} = k\rho\sin\theta - 2\pi A'_{\rm c}\frac{\sin^3 \theta}{\sin^3 \theta_{0}} e^{-i(\phi_{\rm c}-\phi)}= |{\cal Z}|\,e^{i\alpha}.
\end{equation}

Most numerical examples discussed in this paper involve the trap stiffness rather than the force itself. Except in the case of coma, we compute the stiffness by first taking
 the spatial derivative (usually with respect to $\rho$) of the partial-wave series for the relevant force component in order to obtain 
the partial-wave series for the trap stiffness itself, which is then numerically evaluated.

 \subsection{Numerical results}

In all numerical examples discussed in this section, we take  a
typical setup often used in quantitative applications. 
We consider a
 polystyrene (refractive index $1.57$) microsphere of radius $a=0.26\,\mu{\rm m}$ immersed in water (index $n_{\rm w}=1.32$)
trapped by a Nd:YAG laser beam (wavelength $\lambda = 1.064\,\mu{\rm m},$ waist $w_0=4.2\,\mbox{mm}$). 
The beam is focused by an oil-immersion (glass index $n=1.51$) 
objective of   $\mbox{NA} = 1.4$ and entrance aperture radius $R_{0}=3.5$ mm.

The formalism presented in  Sec.~\ref{general}
allows one to consider the joint effects of astigmatism, coma and spherical aberration. We begin by considering	
 each primary aberration  separately
in order to grasp their physical effects on the optical force field. 
We start with the simplest one:  spherical aberration.

 \subsubsection{Joint interface and system spherical aberration}
 
 In this sub-section, we assume that the optical setup contains only spherical aberration: $A'_{\rm a}=
 A'_{\rm c}=0.$
 In order to control the amount of interface spherical aberration,	
 we need to evaluate the
 distance $L$ between the paraxial focus and the glass slide [see  Eq.~(\ref{aberration_interface})], which is not directly accessible in our calibration experiments. 
 Experimentally, we start from the 
 configuration in which the bead touches
 the glass slide at the bottom of the sample chamber and then displace the inverted objective upward by a known distance $d.$ Hence the paraxial focus is 
 displaced vertically from its initial position by a distance $Nd.$
 We mimic this experimental procedure in the following way. We first compute the initial distance between  the paraxial focus and the glass slide $L_0$
  by using the  condition that the bead is initially at equilibrium just touching the glass slide. Then we take $L=L_0+Nd.$
 
 We consider the joint effect of the 	
 Êinterface and system  spherical aberrations on the trap stiffness. We first compute the equilibrium position
 $z_{\rm eq}$ by solving the implicit equation
$Q_ {z} (\rho = 0,  z_{\rm eq}) = 0$. 
The resulting position  is slightly above the diffraction focus because of radiation pressure, and below the paraxial focus for $A'_{\rm sa}\le 0$
(the ratio between the displacement of the equilibrium position and $d$ is usually known as `effective focal shift' \cite{Neuman2005}). 
We  derive 
partial-wave series for the dimensionless force derivatives from the results given in  Appendix A and then take $\rho=0,$ $z=z_{\rm eq}$ to calculate the axial stiffness per unit power; 
 similarly for the transverse stiffness $k_{\rho}.$

\begin{figure}
\begin{center}
 \includegraphics[width=8.5cm]{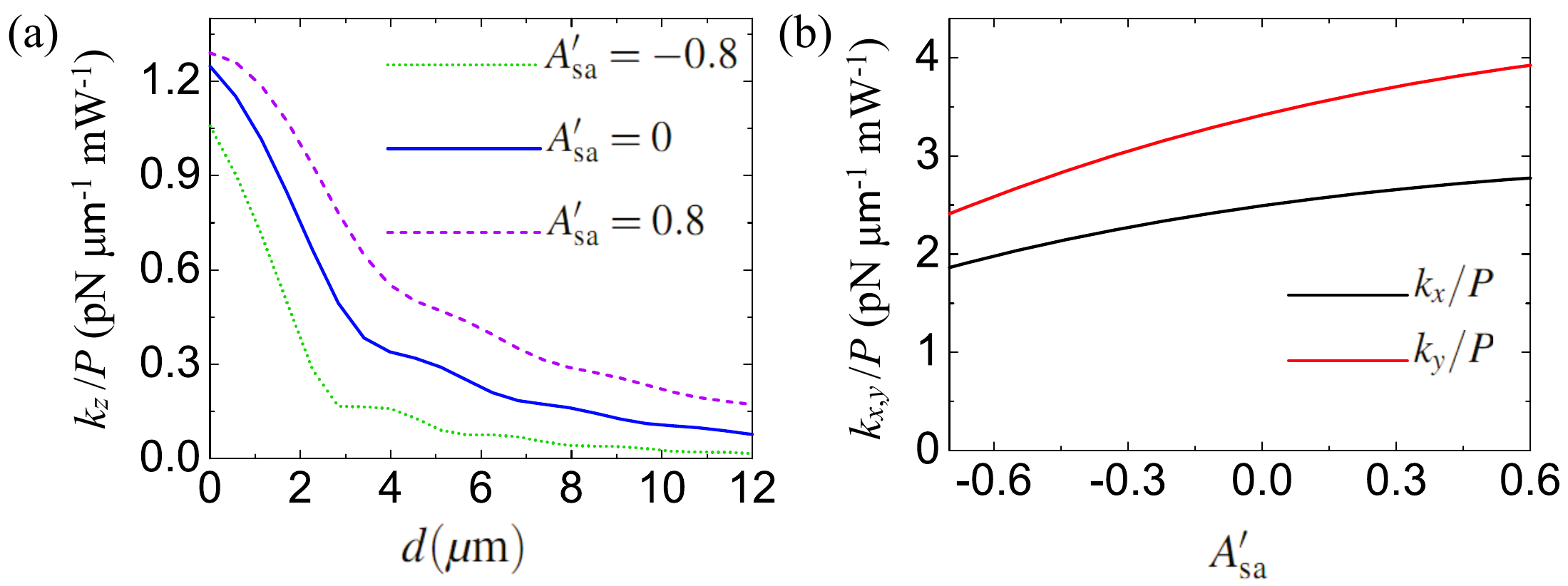}\end{center}
\caption{\label{f1} (Color online) Trap stiffness dependence on system spherical aberration.
 { a)} Axial stiffness per unit power as a  function of the objective upward displacement $d,$ for different values of the 
system spherical aberration amplitude $A'_{\rm sa}.$
{ b)} Transverse  stiffnesses  per unit power
as a function of  $A'_{\rm sa}$ for a fixed objective displacement $ d =3\, \mu{\rm m}.$ The stiffness is larger along the direction perpendicular ($y$ direction)
to the laser beam polarization at the objective entrance. 
}
\end{figure}

In Fig.~1(a), we plot the axial trap stiffness $k_z$ as a function of the objective upward displacement $d,$ for different values of the 
system spherical aberration amplitude $A'_{\rm sa}.$ The solid line, representing the case with only interface spherical aberration, is very similar to the result 
found in Ref.~\cite{Neuman2005}. 
As expected, 
increasing the focal height with respect to the glass slide degrades the focal region,   leading to a severe axial stiffness reduction. 
Since the interface spherical aberration is negative (i.e. the real wavefront is ahead of the ideal spherical reference wavefront), a positive 
 $A'_{\rm sa}$ leads to a partial compensation of the interface effect, as shown in Fig.~1(a), whereas a negative $A'_{\rm sa}$ enhances the focal region degradation.

The transverse stiffness per unit power
 $k_{\rho}$ is less sensitive but also decreases with the trapping height. Here again a positive system spherical aberration partially 
compensates the effect of the interface one. In Fig.~1b, we plot $k_{\rho}$ as a function of  $A'_{\rm sa}$ taking $d= 3\, \mu{\rm m}. $
The stiffness is larger along the direction perpendicular to the incident polarization 
($\phi=\pi/2$ corresponding to $k_y$) because the electric energy density gradient is larger 
along this direction \cite{RichardsWolf59}.

 \subsubsection{Coma}

\begin{figure}
\begin{center}
 \includegraphics[width=8.8cm]{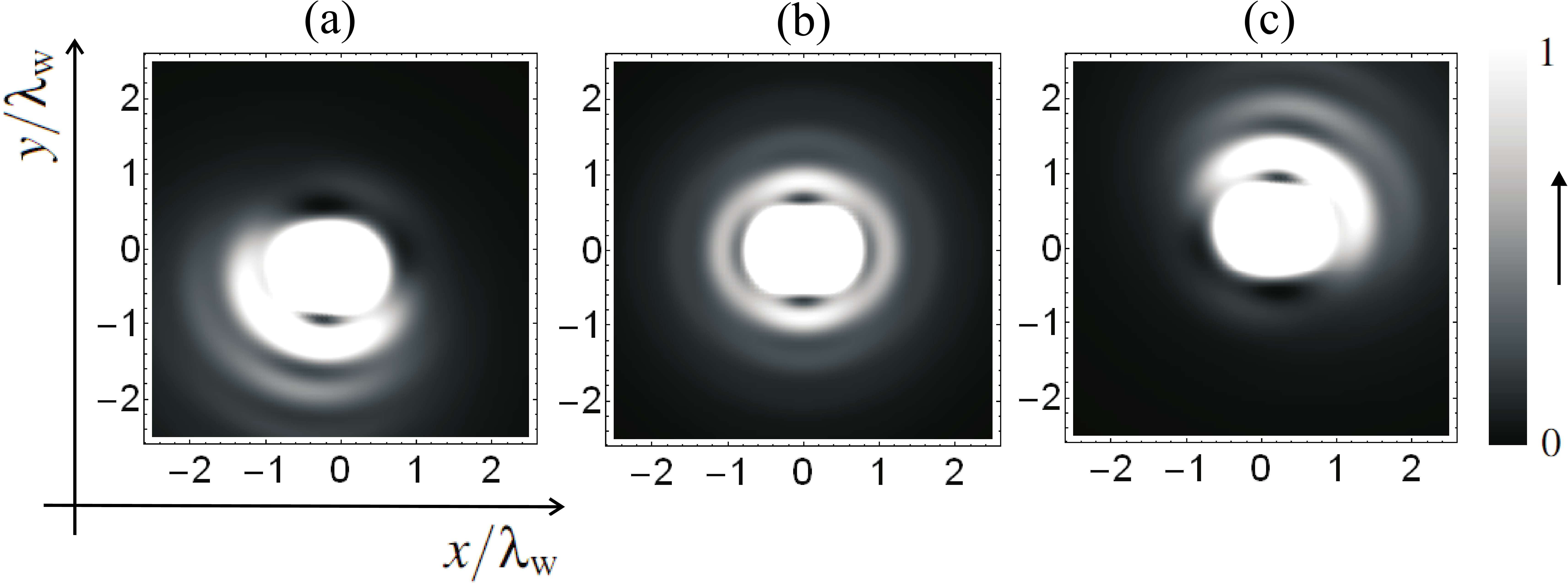}
 \end{center}
\caption{\label{spot_coma} 
Theoretical relative
electric energy density $E^2/E^2_{\rm max}$ on the plane  $z=z_{\rm eq}$
in the presence of coma ($\lambda_{\rm w}=\lambda/n_{\rm w}$ is the wavelength in the sample medium).
 We take $\phi_{\rm c}=\pi/3$ and (a) $A'
 _{\rm c}= -0.93,$ (b) $A'_{\rm c}=0$ and (c)  $A'_{\rm c}= 0.93.$
Note that the non-paraxial coma-free
 focused spot (b) is elongated along the polarization direction $x$ of the laser beam at the objective entrance port \cite{RichardsWolf59}.
}
\end{figure}

\begin{figure}[tbph]
\begin{center}
 \includegraphics[width=8.5cm]{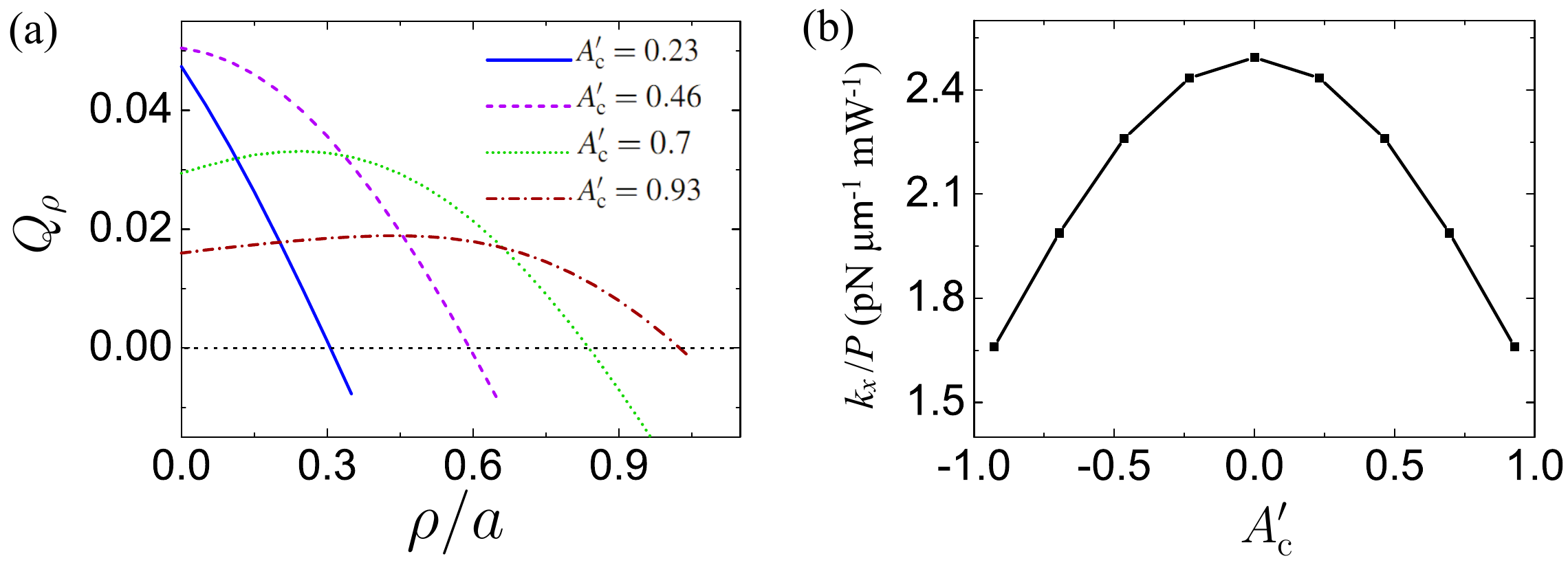}
 \end{center}
\caption{\label{f2} (Color online)  Optical trap with coma. 
(a) Dimensionless radial force $Q_{\rho}$ as 
a function of the cylindrical coordinate $\rho$ along the coma axis $\phi=\phi_{\rm c}.$ The point of equilibrium is off axis.
{ (b)} Transverse trap stiffness per unit power $k_x/P$ along the direction parallel to the incident polarization as a function of  the coma amplitude.}
\end{figure}

When we add coma 
to our setup, the equilibrium position is no longer along the $z$-axis, because
the point of maximum energy density is displaced away from the axis
along  
the direction set by the coma axial direction $\phi=\phi_{\rm c}$ on the $xy$ plane.  
This is illustrated in Figs.~\ref{spot_coma}(a)  and  ~\ref{spot_coma}(c), where we plot the  electric energy density divided by its maximum value
$E^2/E^2_{\rm max}$ 
 at the plane $z=z_{\rm eq}$ corresponding to the axial equilibrium position
 ($E^2=|{\bf E}|^2$ is the electric field square modulus). We also show the spot with zero coma for comparison [\ref{spot_coma}(b)]. 
For all numerical examples presented in Figs.~\ref{spot_coma} and \ref{f2}, 
we take the coma axial direction at $\phi_{\rm c}=\pi/3$ and fix the distance between the paraxial focus and the glass slide to be
$L= 2.9\, \mu{\rm m}.$

We find that the equilibrium position also lies along the coma axis in general (and not only in the Rayleigh regime), regardless of the polarization direction at 
the objective entrance port. 
 In order to determine the full equilibrium position, we first find the coordinate  $z=\bar{z}(\rho)$ yielding axial equilibrium 
as we change the lateral position $\rho$ by solving the implicit equation
\begin {equation}
Q_ {z} (\rho,\phi=\phi_{\rm c},\bar{z}(\rho)) = 0. \label{root}
\end {equation}
We then plot $Q_{\rho}(\rho,\phi=\phi_{\rm c},\bar{z}(\rho))$ as a function of $\rho$ for   different values for the coma amplitude $A_{\rm c}$ in  Fig.~\ref{f2}a.
The distance $\rho_{\rm eq}$ between the equilibrium position and the $z$-axis is given by the intersection between the different curves and the horizontal dashed line 
$Q_{\rho}=0.$ Fig.~\ref{f2}a shows that the equilibrium point is displaced away from the $z$-axis as we increase the coma amplitude, as expected. Moreover,
the figure shows that the equilibrium point is radially stable. By analyzing the dimensionless force components $Q_z$ and $Q_{\phi},$ we find that the equilibrium point is also
stable with respect to axial and tangential displacements. 

As in the coma-free simulations presented  in Ref.~\cite{Mazolli03}, Fig.~\ref{f2}a simulates experiments where a transverse Stokes drag force
$F_{\rm Stokes}$ is applied to the 
trapped microsphere, provided that the Stokes force is parallel to the coma axis. In this case, the new radial equilibrium position can be read
from Fig.~\ref{f2} by taking the value of $\rho$ corresponding to $Q_{\rho}=-cF_{\rm Stokes}/(n_{\rm w}P).$
Note that each value of $\rho$ corresponds to a different axial coordinate ${\bar z}(\rho)$
defined by (\ref{root}), 
 for the microsphere is also displaced along the axial direction when applying the lateral Stokes force \cite{Ashkin1992}
 as demonstrated in 
  Ref.~\cite{Merenda2006}.
 
The Stokes calibration provides perhaps the simplest method for measuring the transverse trap stiffness. 
The radial stiffness $k_{\rho}$ corresponds to the slopes  shown 
in Fig.~\ref{f2}a at $\rho=\rho_{\rm eq}.$ It is already clear from this figure that $k_{\rho}$ decreases with increasing coma amplitude.

It is more common, however, to measure the transverse stiffnesses parallel ($k_x$) or perpendicular ($k_y$) to the polarization axis.  
We calculate $k_x$ for a focused beam with coma from the numerical evaluation of the slope of $Q_x$ in the neighborhood of the point of 
equilibrium. 
 In Fig.~\ref{f2}b,
we plot  $k_{x}$ per unit power as a function of $A_{\rm c}'$ showing that the stiffness reduction does not depend on its sign. 
This symmetry also follows from
 (\ref{funcaoaberracaoum}):
  changing the sign of 
$A_{\rm c}'$ is equivalent to shifting $\phi_{\rm c}\rightarrow \phi_{\rm c}+\pi,$ which amounts to rotating the energy density profile by $\pi,$
as illustrated by Figs.~\ref{spot_coma}a and \ref{spot_coma}c.
 The  
 equilibrium position is then displaced along the opposite direction but the stiffness remains the same. 
These results are in qualitative agreement with the experimental data  presented in Ref.~\cite{Roichman06}.

 \subsubsection{Astigmatism}
 The phase correction corresponding to astigmatism, on the other hand, has a different symmetry property under 
 the change of sign of its amplitude, so that the stiffness is not an even function of $A'_{\rm a}.$
 According to (\ref{funcaoaberracaoum}), 
  when $\phi_{\rm a}\rightarrow \phi_{\rm a}+\pi/2$ the astigmatism phase correction changes sign and yields a residual proportional to $\rho^2,$
 which corresponds to curvature of field.  The latter produces essentially a displacement of the energy density
  profile along the $z$-axis~\cite{Born&Wolf}, with a negligible effect on  stiffness. 
  The transformation $A_{\rm a}'\rightarrow - A_{\rm a}'$  is therefore
   approximately equivalent  to rotating the astigmatism axis by $\pi/2$ 
   \cite{footnote1}.

\begin{figure}[tbph]
\begin{center}
 \includegraphics[width=8.8cm]{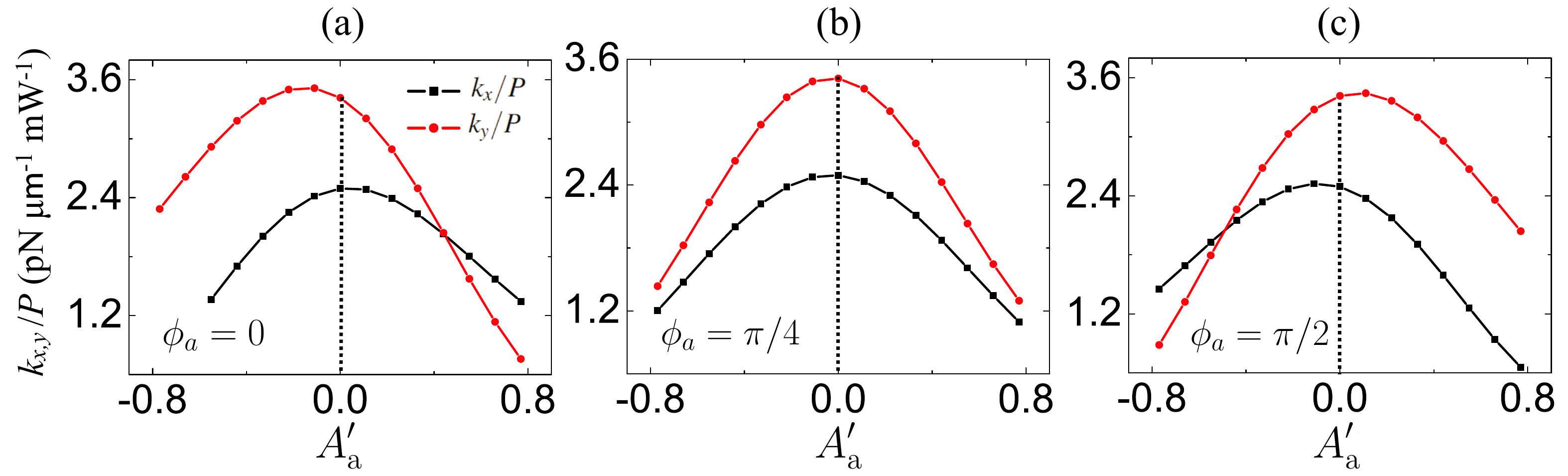}
 \end{center}
 \caption{\label{f3} (Color online) 
Transverse stiffnesses per unit power $k_x/P$ and $k_y/P$ as functions of the astigmatism amplitude $A'_{\rm a}$ for 
 axis orientations (a) $\phi_{\rm a}=0$; (b) $\phi_{\rm a}=\pi/4$ and (c) $\phi_{\rm a}=\pi/2.$ The vertical dashed lines indicate the values in the stigmatic case. 
 The $x$ axis ($\phi=0$) corresponds to the 
 trapping laser beam polarization at the objective entrance.}
\end{figure}

\begin{figure}[tbph]
\begin{center}
 \includegraphics[width=8.7cm]{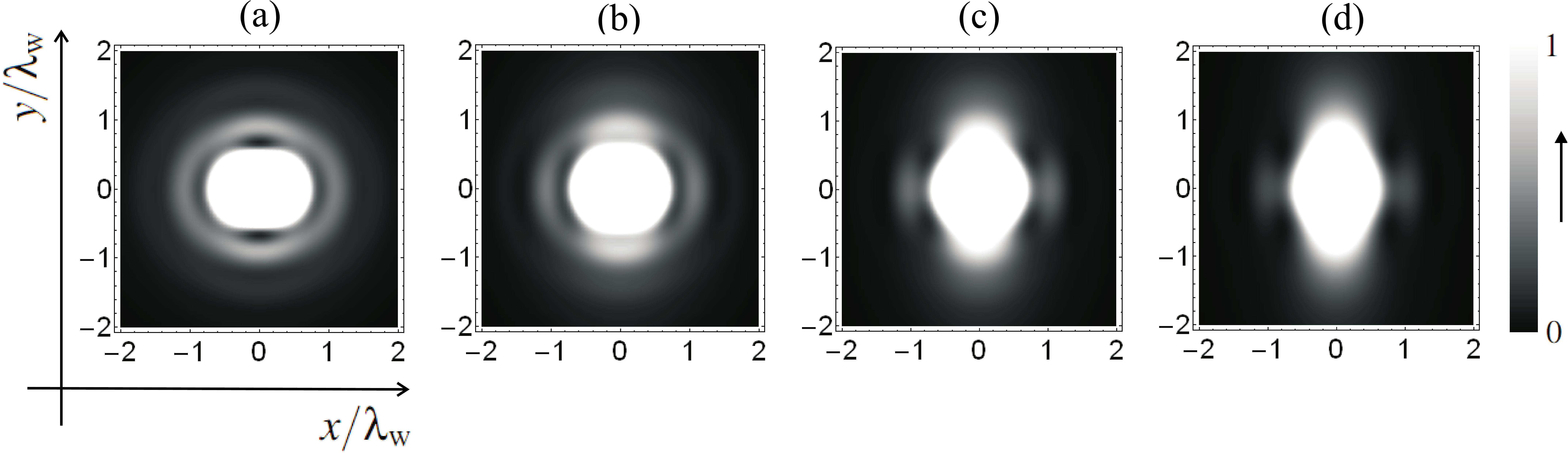}
 \end{center}
\caption{\label{spot_astigmatism} 
Theoretical relative electric energy density $E^2/E^2_{\rm max}$ on the plane  $z=z_{\rm eq}$
in the presence of astigmatism (same conventions as Fig.~\ref{spot_coma}).
 We take $\phi_{\rm a}=0$ and (a) $A'_{\rm a}= 0,$ (b) $A'_{\rm a}=0.22,$  (c)  $A'_{\rm a}=  0.44$ and (d) $A'_{\rm a}= 0.66.$
Note that spot (a) is 
identical to the spot shown in Fig.~\ref{spot_coma}(b). 
}
\end{figure}

  This is verified by the numerical calculations presented in Fig.~\ref{f3}, where 
  we plot the transverse stiffnesses per unit power parallel ($k_x/P$) and perpendicular ($k_y/P$) to the incident polarization as functions of $A_{\rm a}'.$ 
 We take a fixed objective displacement $d = 3\, \mu{\rm m}$ and the astigmatism axis orientations $\phi_{\rm a}=0$ (\ref{f3}a), $\phi_{\rm a}=\pi/4$ (\ref{f3}b), 
 and $\phi_{\rm a}=\pi/2$ (\ref{f3}c). 
 The values for $A'_{\rm a}=0,$ indicated by vertical dashed lines, are of course the same for the three plots and show that the stiffness is larger along the direction perpendicular to the incident polarization as expected, since the
energy density spot at the focal plane in the non-paraxial regime is elongated along the incident polarization direction $x$
in the stigmatic case \cite{RichardsWolf59}, as shown by Fig.~\ref{spot_astigmatism}a.  

By changing the spot shape on the $xy$ plane, astigmatism produces
 a strong effect on the transverse stiffnesses and in particular on their relatives values. 
 The relative electric energy density at the plane 
 $z=z_{\rm eq}$
 is shown in Fig.~\ref{spot_astigmatism}, with
  the astigmatism axis at $\phi_{\rm a}=0.$  
 In order to understand the results shown in Figs.~\ref{f3} and \ref{spot_astigmatism}, we have to bear in mind that  
radiation pressure pushes the equilibrium point to a plane above the diffraction focus (circle of least confusion). For that reason, 
when taking $\phi_{\rm a}=0$ (Fig.~\ref{f3}a)  the spot on the equilibrium plane  $z=z_{\rm eq}$ gradually becomes more elongated 
along the $y$ axis as we increase $A_{\rm a}',$
as illustrated by Fig.~\ref{spot_astigmatism}. 
  As a consequence,  $k_y$ decreases very fast, whereas $k_x$ is initially constant and then starts to decrease as well, since larger values of astigmatism will ultimately degrade the energy density gradient also along $x.$ 
For  $A'_{\rm a}=0.44$, astigmatism yields an exact cancelation of the non-paraxial effect on the spot shape and then we have $k_x=k_y.$
Beyond that point, astigmatism dominates and the spot becomes more elongated along the $y$ direction, yielding $k_x>k_y.$

On the other hand, the gradual introduction of a 
negative astigmatism ($A'_{\rm a}<0$) makes the spot still more elongated along the polarization direction $x,$ reinforcing the gradient along $y$ for moderate 
values of $A'_{\rm a}.$ Thus, $k_y$ is slightly increased by the introduction of a small negative astigmatism as shown by Fig.~\ref{f3}a. Larger values 
of $A'_{\rm a}$ will ultimately degrade both $k_x$ and $k_y.$

For $\phi_{\rm a}=\pi/4,$
 (Fig.~\ref{f3}b),  $k_x$ and $k_y$ become approximately even functions of $A_{\rm a}'$ as expected, since changing the sign of the amplitude is
 equivalent to rotating the axis by $\pi/2,$ apart from a very small contribution from curvature of field.  This symmetry is also apparent when comparing the results 
 for $\phi_{\rm a}=0$ (Fig.~\ref{f3}a) with those for $\phi_{\rm a}=\pi/2$ (Fig.~\ref{f3}c).

By comparing figures \ref{f1}b, \ref{f2}b and \ref{f3}, we conclude that astigmatism is the primary aberration yielding the strongest effect on the 
 transverse stiffnesses $k_x$ and $k_y,$ which are very sensitive to the amplitude $A_{\rm a}',$ again in agreement with the experimental results of Ref.~\cite{Roichman06}.
 Fig.~\ref{f3} 
 shows that the astigmatism axis orientation is also extremely important. This overall message will be of great value in the next two sections, where we undertake the 
task of performing an absolute calibration of stiffness.

 \section{Measuring the astigmatism parameters}
 
\subsection{Experimental procedures}

\begin{figure}[tbph]
\begin{center}
 \includegraphics[width=8.7cm]{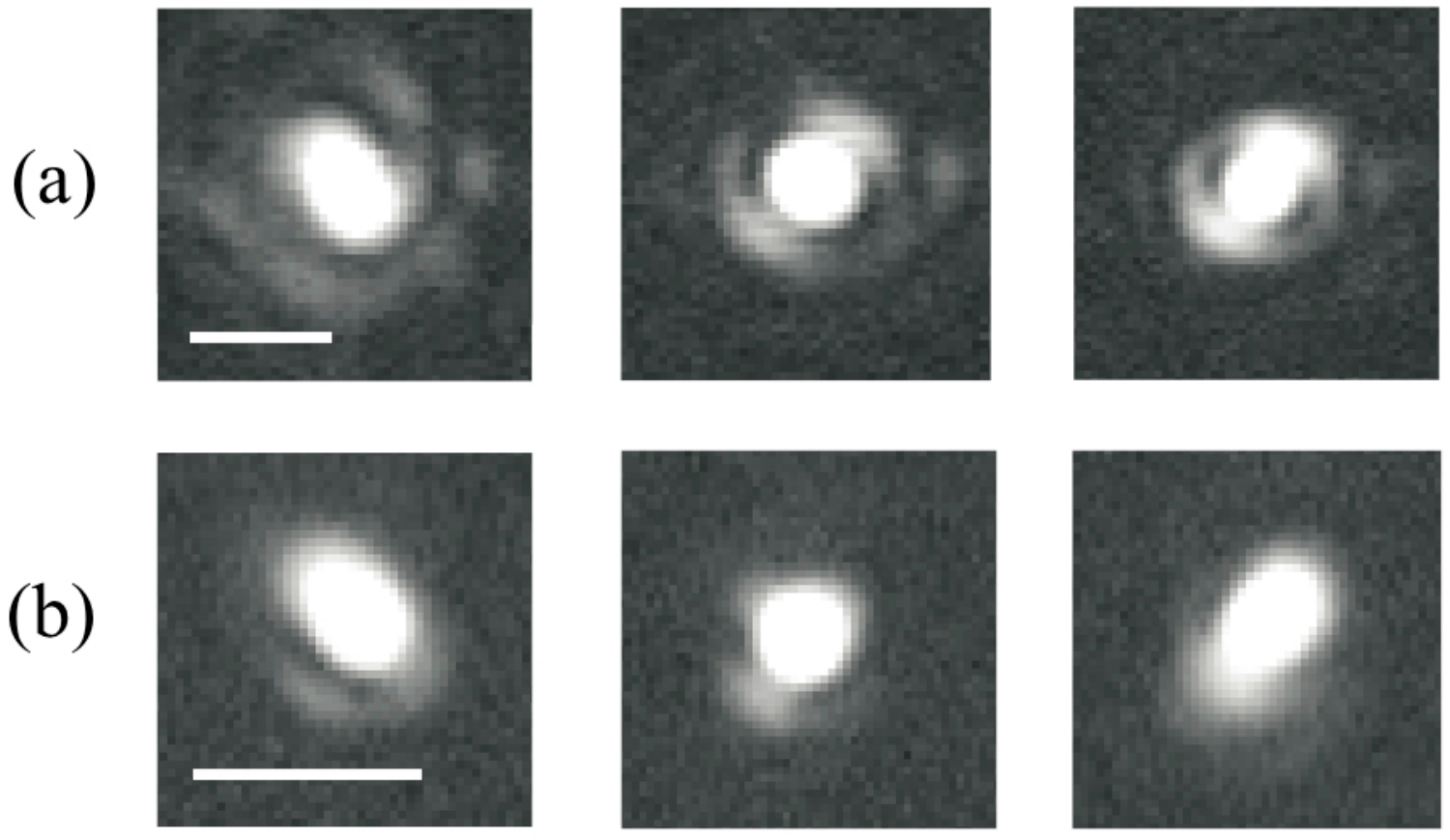}
 \end{center}
\caption{\label{f5}
 From left to right, 8-bit laser spot images below, at and above the circle of least confusion. Two different objectives were employed:  (a) Plan Apo, NA 1.4, 60X; and 
(b) Plan Fluor, NA 0.3, 10X. Scale bars (a) $1\,\mu{\rm m},$  (b) $10\,\mu{\rm m}.$}
\end{figure}

In this section, we present the diagnostic procedures employed for the characterization of optical aberrations present in our 
typical OT setup. Images of  the focused laser spot at different planes across the focal region, shown in Fig.~\ref{f5}, have the elongated form 
typical of 
astigmatism (see Fig.~\ref{spot_astigmatism} for theoretical astigmatic spots). 
They do not show the characteristic shape of coma (see Fig.~\ref{spot_coma}),  which we disregard from now on. 
As discussed in
the previous section, 
 the transverse trap stiffness is extremely sensitive to the astigmatism parameters $A'_{\rm a}$ and $\phi_{\rm a}$
when trapping small spheres. 
 Hence a careful characterization of both astigmatism parameters is essential for undertaking a blind theory-experiment comparison.

\begin{figure}[tbph]
\begin{center}
 \includegraphics[width=8.7cm]{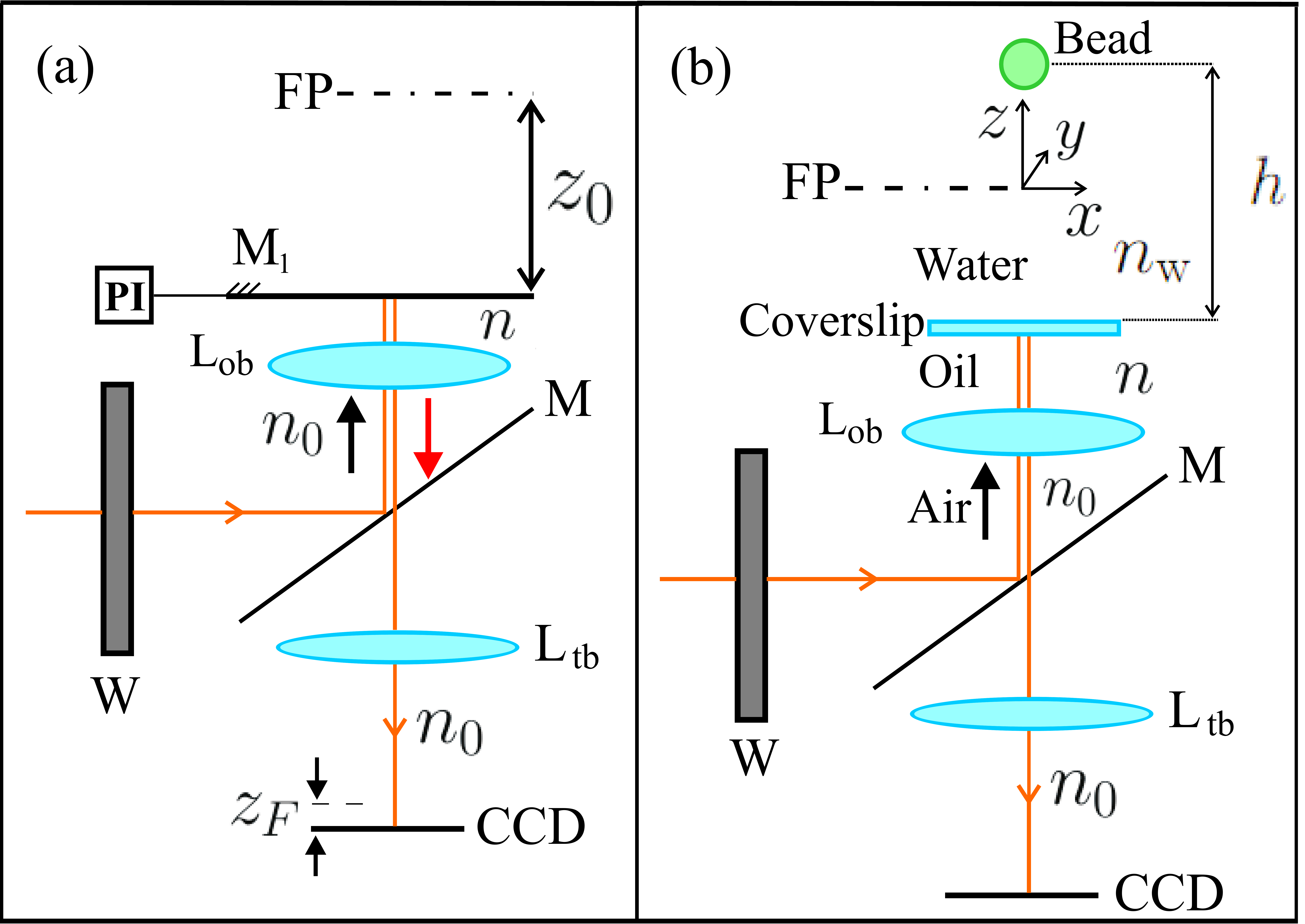}
 \end{center}
\caption{\label{f4} (Color online) 
Schematic representation of the experimental setup for (a)  characterization of astigmatism; and 
(b)  measurement of the transverse trap stiffness. W = waveplate, M = dichroic mirror; ${\rm L}_{\rm ob}=$ objective lens; 
${\rm M}_1 =$ laser spot reflecting mirror; PI = piezoelectric controller;  FP = objective focal plane; ${\rm L}_{\rm tb} =$ tube lens; CCD = recording camera.}
\end{figure}

 Our method is based on the quantitative analysis of the 
  images of the focused laser spot reflected by a plane mirror placed near the focal region, as represented in Fig. \ref{f4}a.  
The collimated $\mbox{TEM}_{00}$ Nd:YAG laser  beam 
(wavelength $\lambda= 1.064\, \mu{\rm  m},$ waist $w_0=4.2\,\mbox{mm}$)  is transmitted through a waveplate W (quarter or half wavelength) that allows to
 control its polarization
at the back entrance of a
 Nikon Eclipse TE300   oil-immersion inverted  microscope (Nikon, Melville, NY). 
 After partial reflection  by the dichroic mirror M (80\% reflectivity), the laser beam propagates in air 
 (refractive index $n_0$) and  reaches
 the objective lens ${\rm L}_{\rm ob}$ (Nikon PLAN APO,  NA 1.4,  60X, 
   aperture radius $R_{0} = 5.0\pm 0.1\,{\rm mm}$ and
  focal distance $f=0.5\,{\rm cm}$)  that focuses the laser beam into a spot localized at the objective focal plane FP  in the immersion oil medium of refractive index $n$. 
  The mirror  $\rm M_1$ (99\% reflectivity)  at position $z_0$   reflects the laser beam back towards  the objective. On its way back a small fraction of the power is transmitted by the mirror M and the spot image is conjugated  by the tube lens ${\rm L}_{\rm tb}$ (focal distance $f_{\rm tb}=20\,{\rm cm}$) onto a
  CCD (charge-coupled device)  camera, which records the  defocused spot image. 
  We employ the piezoelectric nanopositioning system PI (Digital Piezo Controller E-710, Physik Instrumente, Germany)  to move the
  mirror $\rm M_1$  across the focal region with controlled velocity $V=100\,{\rm nm/s}.$   Images of the entire process are recorded
   using a LG7 frame grabber (Scion, USA) connected to a computer.
  
  Typical images are shown in Fig.~\ref{f5} with (a) the high NA objective used for trapping and (b) a low NA objective.  
  We use (b) to infer the astigmatism phase $\Phi_{\rm s}$
   introduced by the set of lenses and mirrors along the optical train between the laser and the objective entrance port in the actual trapping setup, 
  since the optical aberration introduced by a carefully aligned  low NA objective is negligible. 

 On the other hand,   the images collected with the high NA objective used for trapping contain the information on the astigmatism phase $\Phi_{\rm ob}$ 
   introduced by the objective itself. Since the image is formed after back and forth propagation through the objective, the corresponding total phase 
   is $\Phi_{\rm t}=2\Phi_{\rm ob}+\Phi_{\rm s}.$ In short, we measure $\Phi_{\rm s}$ with the help of the
    low NA objective, and then measure $\Phi_{\rm t}$ with the high NA objective used for trapping. 
By combining the two results, we infer the total  OT
astigmatism phase 
\begin{equation}\label{Phi_OT}
\Phi_{\rm \scriptscriptstyle OT}=\Phi_{\rm ob}+\Phi_{\rm s}
\end{equation}
 for the trapping beam at the sample region, which is the relevant
one for the evaluation of the trapping force using the MDSA+ theory presented in Sec.~2. 

It is simpler to add the different phases in terms of the Zernike polynomials (origin at the diffraction focus) \cite{Born&Wolf}. 
To do this,  we write the astigmatism phase as 
\(
\Phi_{\rm \scriptscriptstyle OT}(\rho,\varphi)=2\pi\,A_{\rm \scriptscriptstyle OT}\,(\rho/R_0)^2\cos[2(\varphi-\phi_{\rm \scriptscriptstyle OT})]
\)
and likewise for $\Phi_{\rm s},$ $\Phi_{\rm t}$ and $\Phi_{\rm ob},$  in terms of the amplitudes  $A_{\rm s},$ $A_{\rm t}$ and  $A_{\rm ob}$  and polar angles
 $\phi_{\rm s},$ $\phi_{\rm t}$ and  $\phi_{\rm ob}.$ The connection with the Seidel formalism employed in Sec.~2 is straightforward: 
 we take $A'_{\rm a}= 2A_{\rm \scriptscriptstyle OT}$ and $\phi_{\rm a}= \phi_{\rm \scriptscriptstyle OT}$ and plug the resulting values
  into the general formalism 
 developed in Sec.~2. 

In order to connect the astigmatism phases to the  images recorded by the CCD represented in Fig.~\ref{f4}a, we extend the non-paraxial formalism for  field propagation developed in~\cite{Novotny01} to  astigmatic  spots.
This allows us to write the electric field after propagation
 through the  optical elements  represented in Fig.~\ref{f4}a  in terms of the astigmatism parameters $A_{\rm t}$ and
 $\phi_{\rm t}$
 (when using the high NA objective  ${\rm L}_{\rm ob}$)  or in terms of $A_{\rm s}$ and $\phi_{\rm s}$ (when  ${\rm L}_{\rm ob}$  is replaced by the low NA objective).
As in~\cite{Novotny01}, we compute the propagated field to lowest order of  $f/f_{\rm tb}.$ In addition, we also
assume that mirror  $\rm M_1$ is a perfect reflector and find the electric field at the point $(\rho_F,\phi_F,z_F)$ in the image space of the tube lens 
 ${\rm L}_{\rm tb}$ (see Sec.~2.1 for the definitions of the field amplitude $E_p$ and the filling factor $\gamma$):
\begin{widetext}
\begin{eqnarray}
\label{campofinaldois}
\mathbf{E}_{\rm \scriptscriptstyle CCD}= -i\frac{E_{p}k_{0}f^2} {f_{\rm tb}}e^{i k_{0}(f_{\rm tb}-z_F)}e^{2ikf}
 \int_{0}^{\theta_{0}}d\theta \sin\theta\cos\theta  e^{-\gamma^2\sin^2\theta}
 e^{2i k z_0 \cos\theta} e^{ i\frac{k_{0}z_{F}}{2}\frac{f^2}{f_{\rm tb}^2}\sin^2\theta}\, \left(g^{(+)}_{1}+
\frac{1}{2}e^{-2i\phi_{F}}g^{(-)}_{1}\right)
\mathbf{\hat x}
\end{eqnarray}
\end{widetext}
The astigmatism parameters are contained in the functions $g^{(\pm)}_{m}(\rho_F,\phi_F,\theta) $ defined in Eq.~(\ref{function_f}) ($m=1$). Here we take 
the coma amplitude to be zero ($A'_{\rm c}=0,$ $\alpha=0$), in addition to ${\cal Z}= k_0\rho_F(f/f_{\rm tb})\sin\theta$ and $A'_{\rm a}=2A_{\rm t},$ $\phi_{\rm a}=\phi_{\rm t}.$   
When considering the low NA setup, we take $A'_{\rm a}=2A_{\rm s},$ $\phi_{\rm a}=\phi_{\rm s}$ and replace $\theta_0$ by the much smaller angular aperture corresponding to $\mbox{NA}=0.3.$

We  measure the energy density variation with the mirror position $z_0$ using the CCD and fit the resulting curve with the help of 
(\ref{campofinaldois}) in order to infer the astigmatism amplitudes, as detailed in the next subsection.

\subsection{Results}

\begin{figure*}[tbph]
\centering
 \includegraphics[width=18cm]{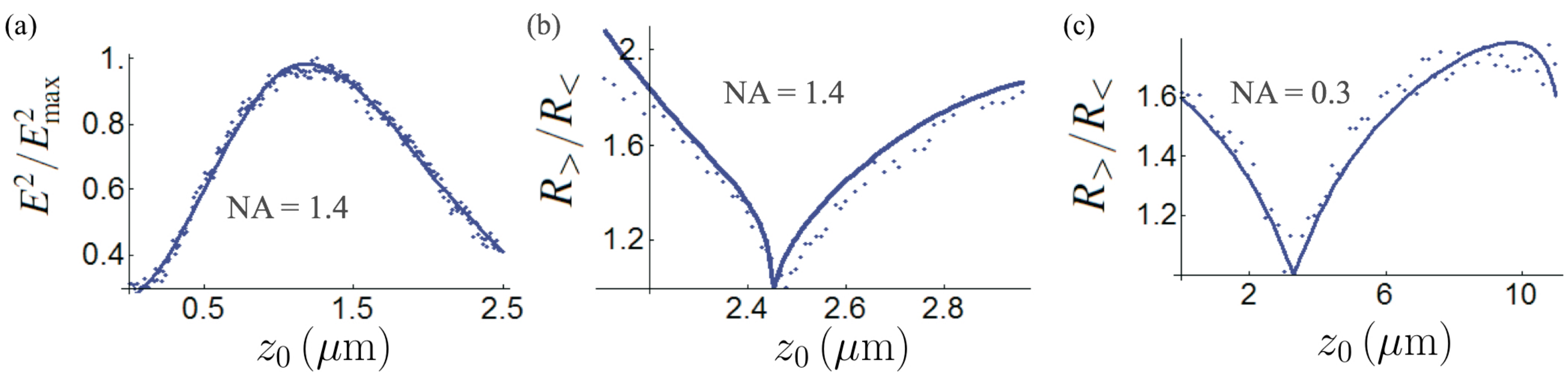}
\caption{\label{f6} (Color online) 
Characterization of astigmatism: experimental data (circles) and curve fit (solid) (a) for the
relative  axial electric energy density $E^2/E^2_{\rm max}$ versus mirror position $z_0$
(NA=1.4 objective);
 for the
ratio of spot radii $R_{>}/R_{<}$ versus $z_0$ with (b) NA=1.4 objective and (c) NA=0.3 objective.  
For the parameters employed in the fits see Tables 1-3.
}
\end{figure*}

In Fig.~\ref{f6}a, we plot
a typical result for 
 the axial ($\rho_F=0$) relative energy density,
$E^2/E^2_{\rm max},$ as a function of $z_0.$ We fit the experimental data by taking the square modulus of (\ref{campofinaldois}). 
In table 1, we show the results for the fitting parameters 
$A_{\rm t},$ $E^2_{\rm max},$ $z_F,$ representing the position of the CCD (see Fig.~\ref{f4}a), and the mirror's position offset $z'$ 
($ z_ {0} \rightarrow z_ {0} - z'$). Each line in table 1 corresponds to a 
 different measurement. 
Since the axial energy density does not depend on the astigmatism orientation axis, we are allowed to combine results for different polarization directions here.

\begin{table}[h]
\begin{center}
\begin{tabular}{|c|c|c|c|c|}
\hline
measurement & $A_{t}$ & $E^2_{\rm max}\,(\mbox{arb. unit})$ & $z_{F}(\mbox{cm})$ & $z'(\mu{\rm m})$ \\
\hline
 1 & 0.95 & 2.8 & 4.9 & -8.5 \\
\hline
2 & 0.98 & 2.7 & 5.4 & -8.8 \\
\hline
3 & 0.99 & 2.7 & 5.5 & -9.6\\
\hline
4 & 0.94 & 2.7 & 5.3 & -9.7\\
\hline
5 & 0.99 & 2.5 & 5.6 & -10.7 \\
\hline
6 & 0.91 & 2.5 & 4.6 & -8.6 \\
\hline
7 & 0.9 & 2.9 & 4.9 & -8.4 \\
\hline
8 & 0.97 & 2.6 & 5.1 & -10.0 \\
\hline
9 & 0.97 & 2.7 & 4.8 & -9.3 \\
\hline
10 & 0.99 & 2.6 & 5.1 & -9.9 \\
\hline
\end{tabular}
\caption{\label{tab1}
Parameters employed for the curve fit 
of the relative axial electric energy density 
(see Fig.~\ref{f6}(a) for a typical example):
 $A_{t}$ (astigmatism amplitude), $E^2_{\rm max}$ (maximum energy density), $z_{F}$ (plane of detection) and $z'$ (offset), for 
 NA = 1.4.}
\end{center}
\vspace{-0.6cm}
\end{table}

\begin{table}[h]
\begin{center}
\begin{tabular}{|c|c|c|c|c|}
\hline
measurement & $A_{t}$ & $E^2_{\rm ctr}\,(\mbox{arb. unit})$ & $z_{F}(\mbox{cm})$ & $z'(\mu\mbox{m})$ \\
\hline
 1 & 0.85 & 7.0 & 5.6 & -7.9 \\
\hline
2 & 0.88 & 8.4 & 5.6 & -8.0 \\
\hline
3 & 0.84 & 9.1 & 5.6 & -8.0\\
\hline
4 & 0.93 & 7.7 & 5.6 & -8.0\\
\hline
5 & 0.82 & 8.4 & 5.4 & -7.6 \\
\hline
6 & 0.84 & 7.7 & 5.7 & -8.2 \\
\hline
7 & 0.83 & 6.9 & 5.6 & -8.0 \\
\hline
\end{tabular}
\caption{\label{tab2}
Parameters employed for the curve fit 
of the ratio $R_{>}/R_{<}$ between 
the major and minor semi-axes of the elliptical contour line  in the $xy$ plane corresponding to a given electric energy density 
 $E^2_{\rm ctr},$
for the  NA=1.4 objective used for trapping
(see Fig.~\ref{f6}(b) for a typical example).
Same conventions as Table 1.
}
\end{center}
\vspace{-0.6cm}
\end{table}

\begin{table}[h]
\begin{center}
\begin{tabular}{|c|c|c|c|c|}
\hline
measurement & $A_{t}$ & $E^2_{\rm ctr}\,(\mbox{arb. unit})$ &   $z'(\mu\mbox{m})$ \\
\hline
 1 & 0.25 & 14.0 & 4.0 \\
\hline
2 & 0.19 & 9.2 & 4.9 \\
\hline
3 & 0.24 & 6.1  & 3.3\\
\hline
4 & 0.22 & 7.0 & 4.2\\
\hline
\end{tabular}
\caption{\label{tab3}
Parameters employed for the curve fit 
of the ratio $R_{>}/R_{<}$ 
for the  NA=0.3 objective  used for measuring the system astigmatism 
(see Fig.~\ref{f6}(c) for a typical example).
Same conventions as Table 2.
}
\end{center}
\vspace{-0.6cm}
\end{table}

 The quality of each fit is extremely sensitive to $A_{\rm t}:$ changing $A_{\rm t}$ by only 5\% leads to a tenfold increase of $\chi^2.$ 
The astigmatism amplitude, averaged out over the 10 measurements shown in Table I, is
 $A_{\rm t}= 0.96\,\pm \, 0.02.$ 

In order to determine the  axis directions $\phi_{\rm t}$ and $\phi_{\rm s},$ we take the elongated spots shown in Fig.~\ref{f5} and fit the 
contour line corresponding to a given value $E^2_{\rm ctr}$ with an ellipse. The resulting directions do not depend on $E^2_{\rm ctr}.$
We find  $\phi_{\rm t}= 57^{\rm o}\pm 3^{\ro}$ and $\phi_{\rm s} = 48 \pm 3^{\ro}$
 for the high  (Fig.~\ref{f5}a) and low  (Fig.~\ref{f5}b) NA objectives, respectively. 

 The ellipses also contain information on the values of the astigmatism amplitudes. 
  We consider the ellipse
   major and minor semi-axes
   $R_{>}$ and $R_{<}$  and plot
   the ratio
 $R_{>}/R_{<}$ 
 versus $z_0$ in Figs. \ref{f6}b (high NA) and \ref{f6}c (low NA). 
 The ratio varies over a much larger distance range  in the second case, as expected in the paraxial regime. 
We fit the resulting experimental data with a theoretical curve calculated from Eq.~(\ref{campofinaldois}).
For the paraxial low NA objective, we can simplify the angular function in the integrand of (\ref{campofinaldois}) and isolate the entire dependence on
$z_0$ and $z_F$ (apart from a trivial phase pre-factor) in terms of the linear combination $-kz_0+(k_0/2)(f/f_{\rm tb})^2 z_F.$ 
Rather than taking  
$z_F$ and the offset $z'$ as independent fitting parameters, we set $z_F=0$
since any finite value of $z_F$ is formally equivalent to a given mirror position offset $z'$ 
 in this case. 
 The results for the fitting parameters are shown in Tables 2 and 3 for the NA 1.4 and NA 0.3 objectives, respectively.  
 
By averaging the values shown in Table 2, we find 
  $A_{\rm t}=0.86\, \pm 0.03,$ close to the value found from the axial energy density distribution. 
  Note that any spherical aberration produced by the objective or by the optical components located between the laser output and 
  the objective entrance would modify the axial energy density but not the ratio $R_{>}/R_{<}.$  Thus, the agreement we have found between the two 
  methods shows that system spherical aberration is negligible in the setup shown in Fig.~\ref{f4}a. This was checked by including spherical aberration in
  Eq.~(\ref{campofinaldois}) and fitting the spherical aberration amplitude $A_{\rm sa}$ using  the axial energy density and the value for $A_{\rm t}$
  found from the ratio $R_{>}/R_{<}.$ The results are distributed around zero with $|A_{\rm sa}|<0.1.$ On the other hand, the interface 
   spherical aberration in the trapping setup  (see Fig.~\ref{f4}b) is very important \cite{NathanPRE} and 
  it is essential to include it 
in the MDSA+ theoretical model. 

We take  $A_{\rm t}=0.92\, \pm 0.04,$ as the overall average combining the two methods. 
From Table~3, we find $A_{\rm s}  =0.23 \pm 0.02$
for the system astigmatism.  
     It is not possible to check this value from the 
   axial energy density variation, which is approximately constant
in the range of distances covered by the PI, as expected in the paraxial regime. 
 We now combine all these values and solve
\begin{eqnarray}
A_{t} \cos 2\phi_{t}&=&A_{s}\cos 2\phi_{s}+2A_{\rm ob}\cos 2\phi_{\rm ob} \\\label{equaum}
A_{t} \sin 2\phi_{t}&=&A_{s}\sin 2\phi_{s}+2A_{\rm ob}\sin 2\phi_{\rm ob}. \label{equadois}
\end{eqnarray} 
to find the objective parameters  $A_{\rm ob} = 0.35 \pm 0.01$ and $\phi_{\rm ob} = (60 \pm 3)^{\ro}$. 
We then combine the objective parameters with $A_{s}$ and $\phi_{s}$ in a similar way [see Eq.~(\ref{Phi_OT})]  and find 
 $A_{\rm OT} = 0.56 \pm 0.03 $ and $\phi_{\rm OT} = 55\pm 3^o$
 (a larger astigmatism amplitude was estimated in a similar setup~\cite{Roichman06}). 
  In the next section, we plug these values into 
 MDSA+ theory and compare the results with the experimental data. 

\section{Transverse Stiffness Calibration}

\subsection{Experimental Procedures}

We validate our proposed absolute calibration by comparison with other known methods \cite{Neuman2004}. 
For testing MDSA theory, 
both Brownian correlations and fluid drag forces were employed as calibration techniques~\cite{NathanPRE},  with comparable results. 
Here we compare MDSA+ with the results obtained by the second approach, with the drag coefficient calculated from Fax\'en's law \cite{Faxen}. 

Our experimental procedures also include the measurement of all input parameters relevant for MDSA+. 
Besides the astigmatism parameters discussed in Sec. III, we also measure the laser beam power and beam waist at the objective entrance port, and the  objective transmittance~\cite{Viana2006}, as described in Ref.~\cite{NathanPRE}. Whenever possible, each input parameter was measured by two different
techniques, checking the results against each other for consistency.

Our OT setup, illustrated by  Fig.~\ref{f4}b, is very similar to the setup for characterization of astigmatism, except for the replacement of
 mirror $\mbox{M}_1$ 
 by a glass coverslip at the bottom of our sample chamber containing polystyrene microspheres (Polysciences, Warrington, PA),  
 diluted to $1 \mu {\rm l}$ of stock solution $10\% v/v$ in $10\, \mbox{ml}$ of water.
 In order to determine the amount of spherical aberration introduced by the glass-water planar interface (see Sec.~2.B.1 for details), 
 we first move down the inverted objective until
 the trapped bead just touches the bottom of the sample chamber. Then we displace the objective upwards through a controlled  distance $d.$

  Once the height of the equilibrium position is  set, 
 we measure the trap stiffness using Fax\'en's law \cite{Faxen} and videomicroscopy. 
 We set the microscope stage to move laterally with a measured velocity $v$ \cite{footnote_error}  either along the $x$ (polarization)  or $y$ direction,
producing a Stokes drag  force $\beta v$ that displaces the bead off-axis through a distance $\delta \rho$ along the same direction.  
We calculate $\beta$ from Fax\'en's law using the values for the bead radius $a$ and height $h$. 
Each run is recorded with a  LG7 frame grabber (Scion, USA). 
From the digitized images of the trapped bead
we determine $\delta \rho$ as a function of $v.$ 
We employ  values of $v$ small enough to probe only the linear range of the optical force: 
$\beta v = k_{x,y} \delta \rho.$ We check that our data for the lateral displacement is a linear function of $v,$   $\delta\rho=\alpha v,$
 determine the coefficient $\alpha$ and then the transverse stiffness $k_{x,y}= \beta/\alpha$ \cite{SM}.  
When comparing with theory, we take the stiffness per unit power $k_{x,y}/P,$ where the power at the sample region $P$ is derived from the measured 
objective transmittance and  power at the objective entrance port. 

\subsection{Experimental results and comparison with MDSA+}

\begin{figure}[tbph]
\begin{center}
 \includegraphics[width=8.7cm]{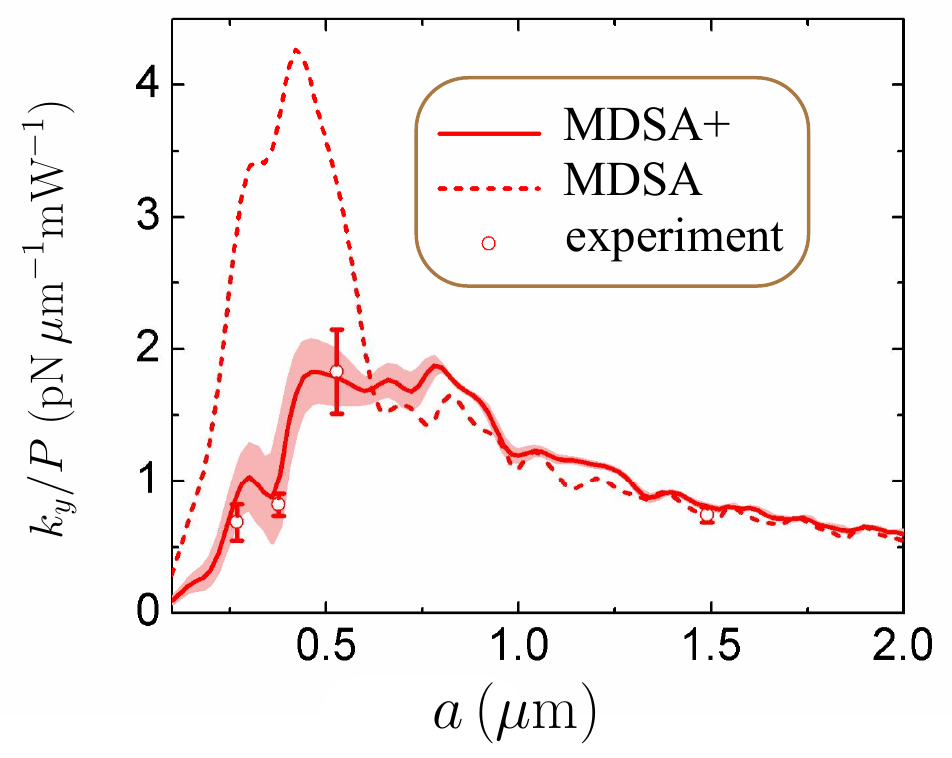}
 \end{center}
\caption{\label{f7} (Color online)  Transverse trap stiffness per unit power  $k_y/P$  
 versus bead radius
   $a$ for  an objective displacement $d=3.0\pm 0.5\,\mu{\rm  m}.$
   No adjustable parameters are employed. 
   Solid line:    
   MDSA+ with the measured astigmatism parameters $A_{\rm OT}=0.56$ and $\phi_{\rm OT}=55^{\ro},$ shaded
    theoretical uncertainty band bounded 
by the curves calculated for 
  $A_{\rm OT}\mp \delta A_{\rm OT},$  $\phi_{\rm OT}\pm \delta\phi_{\rm OT}$ and $d=3.0\mp 0.5\,\mu{\rm m}$
  ($   \delta A_{\rm OT}=0.03$ and $\delta\phi_{\rm OT}=3^{\ro}$), circles:
    experimental results and dashed line: MDSA.   }
\end{figure}

\begin{figure}[htbp]
\begin{center}
 \includegraphics[width=8.7cm]{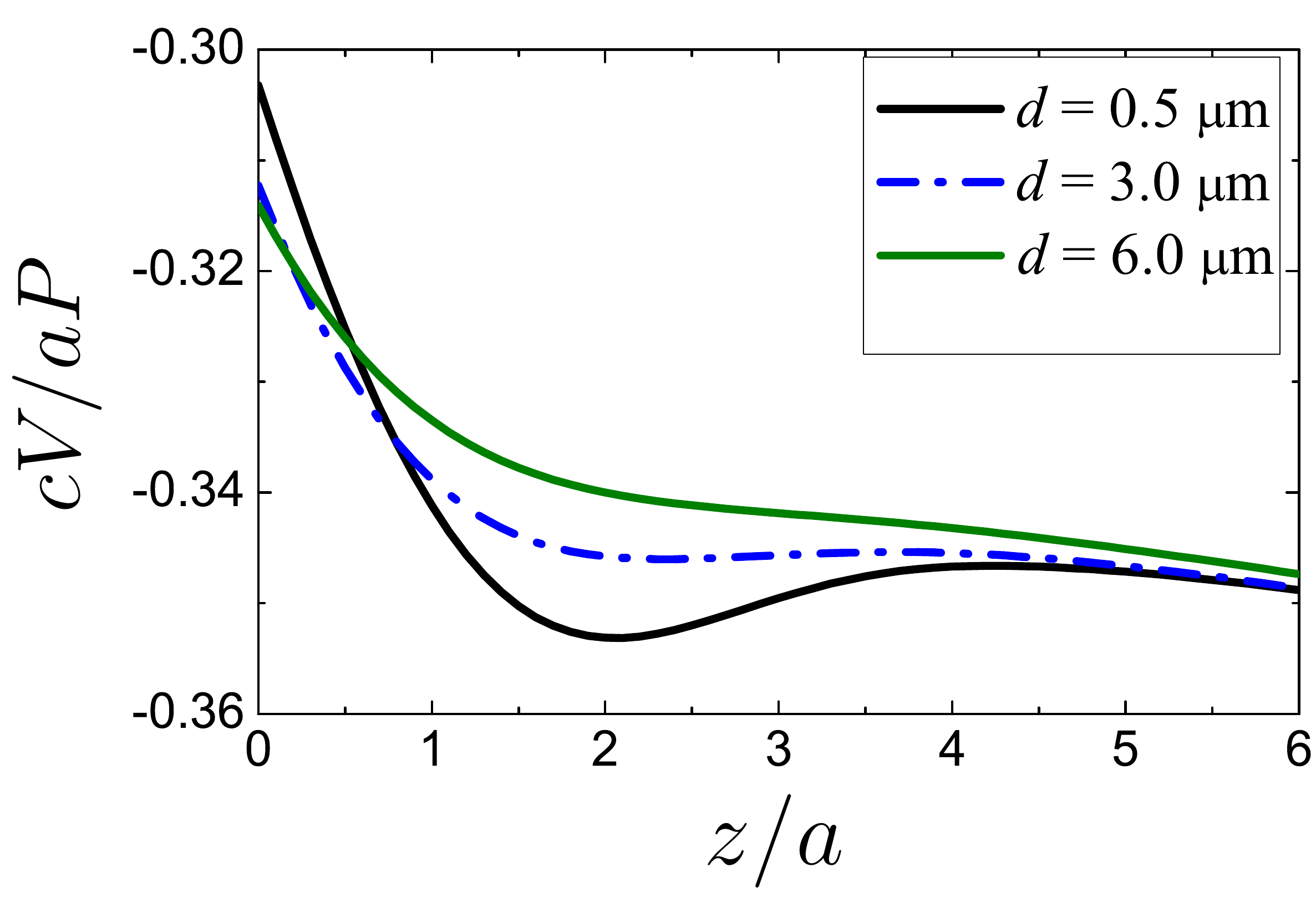}
 \end{center}
\caption{\label{potencial} (Color online)  MDSA+ axial potential $V$ (per unit power divided by $a/c$) versus $z/a$ for 
 $a = 0.376\,\mu{\rm m}.$  The optical potential well becomes shallower as the objective is displaced upwards through  the  distance $d$.
  For $d$ around $3 \mu{\rm m},$ it displays a region of indifferent equilibrium. }
\end{figure}
\begin{figure*}[tbph]
\centering
 \includegraphics[width=18cm]{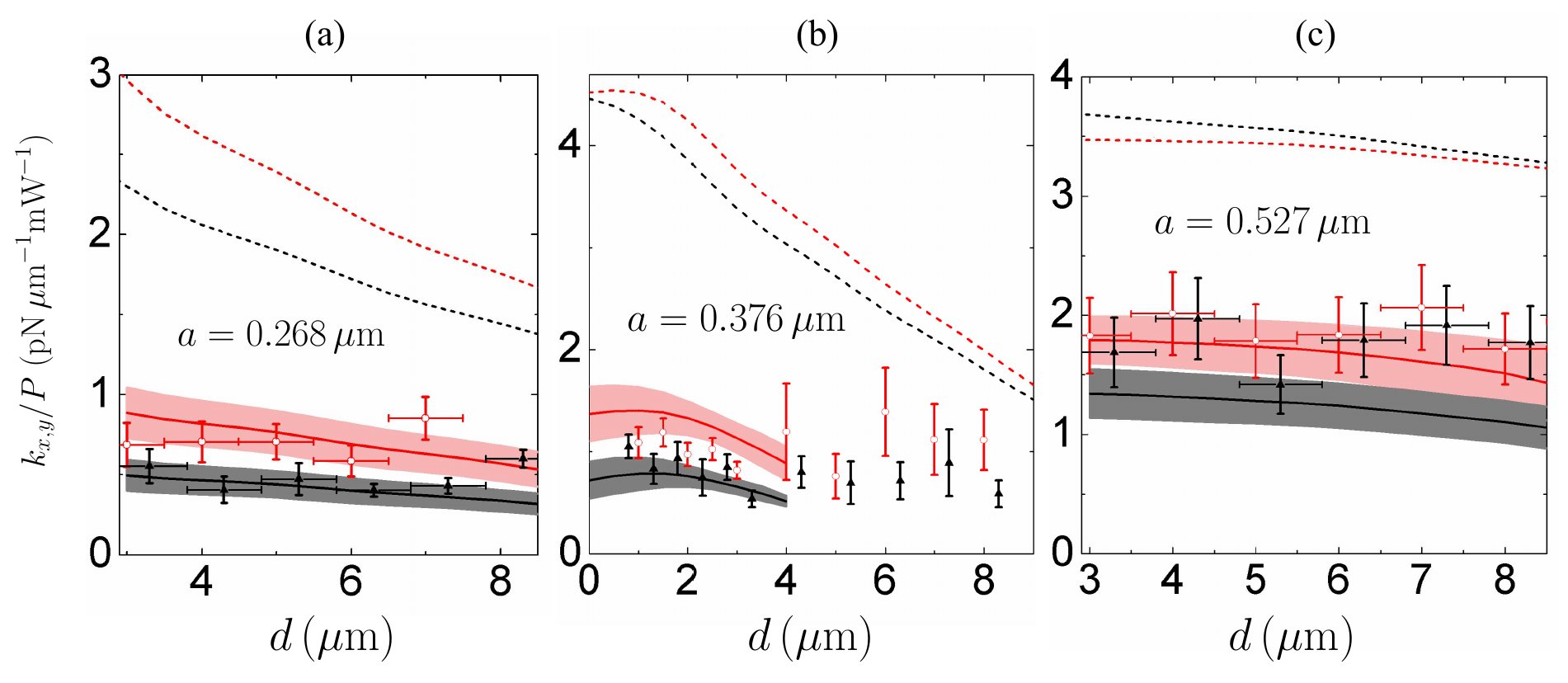}
\caption{\label{variando_d} (Color online)  Transverse trap stiffnesses per unit power versus   
objective vertical displacement $d$ for different microsphere radii:  
(a) $a = 0.268\,\mu{\rm  m},$ 
 (b) $0.376\,\mu{\rm m}$
 and (c) $0.527 \mu{\rm m}$  (same conventions as  Fig.~\ref{f7}).  Black line: $k_x/P;$ red (light gray) line: $k_y/P.$ 
 For clarity, the values of $d$ 
 corresponding to 
 the experimental points for $k_x$ have an offset of $+0.3\,\mu{\rm m},$ and the horizontal error bars (corresponding to $\delta d=0.5\,\mu{\rm m}$) are omitted  in plot (b). }
\end{figure*}

In Fig.~\ref{f7}, we plot the transverse stiffness per unit power $k_y/P$ as a function of bead radius
$a$  for an objective displacement $d = 3.0\pm 0.5\,\mu{\rm m}.$  
All relevant input parameters are 
 determined independently of the stiffness calibration, and no fitting is implemented in the comparison with the experimental results for the trap stiffness discussed in this section. We calculate with the following parameters: beam waist at the objective entrance port $w_0=4.2\,{\rm mm},$ laser wavelength 
 $\lambda=1.064\,\mu{\rm m},$ objective focal length $f=0.5\,{\rm cm},$ polystyrene, water and glass refractive indexes $n_{\rm PS}=1.576,$  $n_{\rm w}=1.332,$
and $n=1.51,$ and semi-aperture angle $\theta_m={\rm sin}^{-1}(n_{\rm w}/n)= 61.9^o.$
 For MDSA+, we also take the measured astigmatism parameters (see Sec. III). 
Fig.~\ref{f7} provides an overall assessment of the stiffness behavior as one sweeps the sphere radius from the Rayleigh $a^3$ increase 
 to the geometrical optics $1/a$  decrease. The MDSA curve (dashed line), corresponding to a 
 stigmatic  beam, develops a peak in the range from $\lambda/4$ to
$\lambda/2,$ at the cross-over between Rayleigh and geometrical optics regimes, in which the stiffness is highly overestimated. Clearly, by including the effect of astigmatism, MDSA+ provides a much better
description of the experimental data in this range. On the other hand,   the effect of astigmatism is reduced
 for larger values of $a,$ as expected, since the details of the energy density distribution are 
averaged out when computing the optical force on a large microsphere. These properties are in qualitative 
agreement with Ref.~\cite{Padgett06}, where the astigmatism correction was
  found to be relevant for a microsphere of radius $0.4\,{\mu}{\rm m}$ but not for large beads. 

The width of the theoretical uncertainty band shown in Fig.~\ref{f7}, 
bounded by
the curves corresponding to parameters $A_{\rm OT}\mp \delta A_{\rm OT}$ and $\phi_{\rm OT}\pm \delta\phi_{\rm OT},$ 
  indicates that the sensitivity to astigmatism is larger for small and moderate bead sizes. 
 More generally, 
the trap becomes more susceptible to perturbations at the crossover between Rayleigh and geometrical optics regimes, 
as exemplified by the effect of astigmatism discussed here. 
This is of considerable practical importance, because this region corresponds to the 
radii most often used in quantitative applications, for which a reliable  transverse stiffness calibration is needed. 

Right at the center of the MDSA peak region shown in Fig.~\ref{f7}, we observe experimentally that 
the trap becomes less stable, particularly for larger trap heights. 
This is well explained by MDSA+. Although the optical force is not conservative \cite{not-conservative}, we can still define
 an effective axial potential as the integral of the axial force component along the $z$ axis, in order to interpret the trap stability in a more intuitive way. We find that there is a window of instability for bead radii in the neighborhood of 
$a=0.376\,\mu{\rm m}$ as we displace the objective upwards. In Fig.~\ref{potencial}, we plot the dimensionless axial potential 
 $cV/(aP)$ versus $z/a$ for 
 $a = 0.376\,\mu{\rm m}.$  The potential well  becomes shallower as $d$ increases and no equilibrium is found for $d=6\, \mu{\rm m}.$
 Experimentally, we find
 a range of approximately indifferent equilibria
 when $d>3\,\mu{\rm m},$
  resulting in a large dispersion of the experimental values. This  translates into the larger experimental error bars shown in Fig.~\ref{variando_d}(b), 
  where we plot $k_x/P$ and $k_y/P$ versus objective displacement $d.$  The axial potential well is also very shallow for $a=0.527\,\mu{\rm m},$
 and the large error bars in  Fig.~\ref{variando_d}(c) are again consistent with this property.

Among the three bead sizes presented in Fig.~\ref{variando_d}, the radius $a = 0.376\,\mu{\rm m}$
right at the instability window 
 is also the one for which we find the largest discrepancy between 
MDSA and the experimental/MDSA+ values. 
In this case, MDSA  overestimates stiffness by a factor larger than 4 for $k_x$ at low heights
 and predicts a steady decrease  as a 
function of $d$ which is not observed experimentally. 
The effect of enhancing the spherical aberration introduced by the glass slide as $d$ increases, which is 
clearly present in the MDSA curves for the two smaller radii  shown in Fig.~\ref{variando_d}, 
becomes less severe  since the energy density gradient is already 
degraded by the presence of astigmatism in  MDSA+. 

Some of the data points shown in Fig.~11(b) correspond
 to bead heights below $1\,\mu{\rm m}.$ Traps very close to the glass slide can be affected by additional perturbations, 
not taken into account in MDSA+, including optical reverberation (multiple light scattering between the glass slide and the microsphere), surface interactions and the contribution of evanescent waves beyond the critical angle. The 
first two effects were carefully probed in Ref.~\cite{Schaffer2007}. For a polystyrene bead of radius $0.264\mu{\rm m},$ an intensity modulation was found for distances below $1\,\mu{\rm m},$ indicating the interference between the trapping beam and the scattered field reflected by the glass slide. This clearly affects the 
equilibrium position, but no effect was found on the transverse stiffness calibration~\cite{Schaffer2007}. However, larger beads at distances below $3\lambda$ from the surface 
might suffer from a stronger reverberation effect, particularly when considering the axial stiffness. 

Fig.~\ref{variando_d}  shows that $k_y$ is larger than $k_x,$ specially for small spheres, which act as local 
probes of the electric energy density profile. In the stigmatic case, the focused spot is elongated along the polarization 
direction \cite{RichardsWolf59}, 
as shown in Fig.~\ref{spot_astigmatism}a, thus leading to a larger gradient along  the $y$ axis. 
This can be reversed by a positive astigmatism when the  axis orientation is smaller than $\pi/4$ 
(see Figs.~\ref{f3} and \ref{spot_astigmatism}). However, in Fig.~\ref{variando_d} we take $\phi_{\rm OT}= 55^{\ro},$ and as consequence the relative difference
between $k_y$ and $k_x$ is actually
 enhanced by astigmatism, specially for the radius $a = 0.376\,\mu{\rm m}.$ In spite of the large
error bars, the experimental data shown in the figure are again consistent with this theoretical prediction.

\section{Conclusion}

Our numerical examples show that even a 
small amount of astigmatism leads to a measurable reduction of the transverse trap stiffness for microsphere radii in the range 
between $\lambda/4$ and $\lambda/2.$ This is of considerable practical importance, as most quantitative applications  rely on 
transverse stiffness calibrations for microspheres precisely in this range.

From a theoretical point of view, this interval of microsphere radii corresponds to the cross-over between the Rayleigh and ray optics regimes. Fig.~\ref{f7}
 provides an overall picture as far as the transverse stiffness is concerned. Right at the crossover, MDSA develops a peak 
(maximum close to $a=0.4\,\mu{\rm m}$  for  $\lambda=1.064\,\mu{\rm m}$), which is severely reduced (and slightly shifted towards larger radii) when astigmatism is included. Therefore, correcting astigmatism, for instance with the help of spatial light modulators \cite{Padgett06, Lopez-Quesada09,Arias13}, might lead in principle to a fourfold increase in the transverse stiffness of our typical OT setup.

Figs.~\ref{f7} and \ref{variando_d} represent a fair sample of the general good agreement between experimental results and MDSA+ that we have found for a variety of bead sizes and trap heights, for circular as well as for linear polarizations, for the transverse stiffness either along $x$ or $y$ directions (the case of circular polarization was briefly reported in~\cite{Dutra2012}). We have also found qualitative agreement with previous measurements of primary
 aberrations effects~\cite{Roichman06,Padgett06}.

With our experimental setup, we have independently measured all parameters needed for the explicit numerical computation of the MDSA+ predictions. In particular, the astigmatism parameters were determined using a simple videomicroscopy method, based on the analysis of the reflected focused spot, that can be easily adapted to any OT setup.  The success of such a blind theory-experiment comparison	demonstrates	that	MDSA+	can	be	used	as	a
practical
calibration	tool, covering the whole range of sizes from the Rayleigh regime to the ray optics one, including the intermediate size interval (peak region) most often employed in applications.

As stated in \cite{NathanPRE} for MDSA, it remains true that MDSA+  does not include the effects of reverberation (multiple light reflections between the bead and the glass slide), and those of evanescent waves beyond the critical angle. Thus, it is advisable when employing it to stay away from the glass slide by at least a couple of wavelengths.  It would be of considerable interest to extend the theory to evanescent wave excitation, so as to provide a theoretical 
description of fluorescence microscopy of single molecules \cite{single_molecules}.

Another promising application is the measurement of surface interactions between
a microsphere and a plane surface~\cite{Schaffer2007} or between 
two trapped microspheres~\cite{Masri2011}.
Absolute OT calibration allows force measurements, currently under way in our laboratory, down to femtonewtons, with the investigation of Casimir forces as a prospect.

In summary, by taking the primary aberrations into account, MDSA+ provides a complete description of the most often employed OT setup when trapping far from the surface. Astigmatism is the primary aberration that produces the largest effect on the transverse stiffness. In our typical setup, it reduces the stiffness by a large factor and, more importantly, it degrades the trap stability for radii close to or slightly smaller than $\lambda/2.$ 
The instability effect could be even more striking when trapping high-refractive index particles in water
 \cite{vanderHorst} or airborne
 aerosol particles \cite{Burnham2006},
 Ê because	
 Êof	
 Êthe	
 Êlarger	
 Êradiation	
 Êpressure	
 Êcontribution	
 Êin	
 Êthese	
 Êcases.
The achievement of absolute calibration signifies that we now have a satisfactory basic understanding of the performance of OT, bringing about the possibilities of improved	design,	fuller	control	and	the	extension	of	the	usual 
 	domain	of applicability of these remarkable instruments, ranging from femtonewtons to nanonewtons.

\acknowledgements

We thank B. Pontes and O. N. Mesquita for discussions. We are indebted to the referee for valuable comments.
This work was supported by the Brazilian agencies CNPq,  FAPERJ  and INCT Fluidos Complexos.

\appendix

\section{Partial-wave series for the dimensionless optical force efficiency}

In this appendix, we write the explicit partial-wave series for the cylindrical components of the dimensionless optical force efficiency $\mathbf{Q}$
defined by eq.~(\ref{defQ}).  

$\mathbf{Q}$ contains two separate contributions:
\(
\mathbf{Q}=\mathbf{Q}_{e}+\mathbf{Q}_{s}.
\)
The extinction contribution $\mathbf{Q}_{e}$ represents
 the rate at which momentum is removed from the focused incident beam. 
 $\mathbf{Q}_{s}=\mathbf{Q}_{s}^{\rm (p)}+\mathbf{Q}_{s}^{\rm (c)}$ represents the negative 
 of the rate at which momentum is carried away by the field scattered by the microsphere (Mie scattering).
 Hence $\mathbf{Q}_{s}$ is quadratic in the scattered field, with
$\mathbf{Q}_{s}^{\rm (p)}$ containing pure electric (magnetic)
multipole contributions, quadratic in the 
 Mie coefficients $a_{j}$ ($b_{j}$) \cite{Bohren&Huffman}, and  $\mathbf{Q}_{s}^{\rm (c)}$ accounting for the  cross terms proportional to $a_{j}b_{j}^*.$ 
 Their cylindrical components are
given by  partial-wave (multipole) sums 
of the form
\[
\sum_{jm\sigma}\equiv \sum_{j=1}^{\infty}\sum_{m=-j}^{j}\sum_{\sigma=\pm 1}.
\]
We find
\begin{widetext}
\begin{eqnarray}
\label{Qsrhop}
Q_{s\rho}^{\rm (p)} &=& \frac{2\gamma^2}{AN}\sum_{jm\sigma}\frac{\sqrt{j(j+2)(j+m+1)(j+m+2)}}{j+1}
 {\rm Im}\biggl\lbrace
(a_{j}a_{j+1}^{*}+b_{j}b_{j+1}^{*})
\Bigl[
G^{(\sigma)}_{j,m}G^{(\sigma)*}_{j+1,m+1}\\
\nonumber &&
+\,G^{(\sigma)}_{j,-m}G^{(\sigma)*}_{j+1,-(m+1)}\Bigr]
+(a_{j}a_{j+1}^{*}-b_{j}b_{j+1}^{*})
e^{2i\sigma\phi}\left[G^{(\sigma)}_{j,m}G^{(-\sigma)*}_{j+1,m+1}+G^{(\sigma)}_{j,-m}G^{(-\sigma)*}_{j+1,-(m+1)}\right]
\biggl\rbrace 
\end{eqnarray}

\begin{eqnarray}
\label{Qsrhoc}
Q_{s\rho}^{\rm (c)}& =  -\frac{4\gamma^2}{AN}\sum_{jm\sigma}
 \frac{(2j+1)}{j(j+1)}\sigma\sqrt{(j-m)(j+m+1)}&\Bigl[{\rm Re}(a_{j}b_{j}^{*})
{\rm Im}( G^{(\sigma)}_{j,m}G^{(\sigma)*}_{j,m+1}) \\
&& +\, {\rm Im}(a_{j}b_{j}^{*}){\rm Re}(e^{2i\sigma\phi}G^{(\sigma)}_{j,m+1}G^{(-\sigma)*}_{j,m})\Bigr]\nonumber
\end{eqnarray}

\begin{eqnarray}
\label{Qszp}
Q_{s z}^{\rm (p)}&=& -\frac{4\gamma^2}{AN}{\rm Re}\sum_{jm\sigma}\frac{\sqrt{j(j+2)(j+m+1)(j-m+1)}}{j+1} \\
&& \times\left[(a_{j}a_{j+1}^{*}+b_{j}b_{j+1}^{*})
G^{(\sigma)}_{j,m}G^{(\sigma)*}_{j+1,m}+(a_{j}a_{j+1}^{*}-b_{j}b_{j+1}^{*})e^{2i\sigma\phi}G^{(\sigma)}_{j,m}G^{(-\sigma)*}_{j+1,m}
\right] \nonumber
\end{eqnarray} 
\begin{eqnarray}
Q_{s z}^{\rm (c)} = -\frac{4\gamma^2}{AN}{\rm Re}\sum_{jm\sigma}
\frac{(2j+1)}{j(j+1)}m\sigma a_{j}b_{j}^{*}\left(\vert G^{(\sigma)}_{j,m}\vert^2-e^{2i\sigma\phi}G^{(\sigma)}_{j,m}G^{(-\sigma)}_{j,m}{}^*\right). \label{Qszc}
\end{eqnarray} 

\end{widetext}
$G^{(\sigma)}_{jm}(\rho,\phi,z)$ are the focused beam multipole coefficients in the case of a circularly polarized beam at 
the objective entrance (helicity $\sigma$),  defined by eq.~(\ref{Gjm}).
The cross terms of the form 
$G^{(\sigma)}_{j,m}G^{(-\sigma)*}_{j',m'}$ in (\ref{Qsrhop})-(\ref{Qszc}) arise from writing the the linearly-polarized field as a superposition of 
$\sigma=\pm 1$ circular polarizations
 and squaring the resulting scattered field when computing the stress tensor. Thus, they are absent in the case of circular polarization discussed in 
 Ref.~\cite{Dutra2012}. 
The 
 filling factor $A$
appearing in (\ref{Qsrhop})-(\ref{Qszc}) represents the fraction of laser bem power transmitted through the objective aperture and the
 glass-slide 
 \cite{footnote2}:
\begin{equation}
A = 16\gamma^2\int_0^{\theta_m} ds\,s\, \exp(-2\gamma^2s^2)\,\frac{\sqrt{(1-s^2)(N^2-s^2)}}{(\sqrt{1-s^2}+\sqrt{N^2-s^2})^2}
\end{equation}

The azimuthal component contributions $Q_{s\phi}^{\rm (p)}$ and $Q_{s\phi}^{\rm (c)}$ are given by expressions similar to 
(\ref{Qsrhop}) and (\ref{Qsrhoc}), respectively. The dimensionless extinction force cylindrical components are given by
\begin{widetext}
\begin{eqnarray}
 \label{Qerho}
Q_{e\rho}=\frac{\gamma^2} {AN}{\rm Im}\sum_{jm\sigma}(2j+1)G^{(\sigma)}_{j,m}
\Bigl[(a_{j}+b_{j})
\left(G^{-,(\sigma)}_{j,m+1}-G^{+,(\sigma)}_{j,m-1}\right)^*+(a_{j}-b_{j})e^{2i\sigma\phi}
\left(G^{-,(-\sigma)}_{j,m+1}-G^{+,(-\sigma)}_{j,m-1}\right)^*\Bigr]
\end{eqnarray}

\begin{eqnarray}
Q_{ez}=\frac{2\gamma^2}{AN}{\rm Re}\sum_{jm\sigma}(2j+1)
G^{(\sigma)}_{j,m}
\left[(a_{j}+b_{j})
G_{j,m}^{C,(\sigma)}{}^*
+(a_{j}-b_{j})e^{2i\sigma\phi}G_{j,m}^{C,(-\sigma)}{}^*\right]
\label{Qez}
\end{eqnarray}
The series representing $Q_{e\phi}(\rho,\phi,z)$ is similar to (\ref{Qerho}).

In addition to the multipole coefficients $G^{(\sigma)}_{j,m}$ defined by Eq.~(\ref{Gjm}), we have also defined

\begin{eqnarray}
G_{j,m}^{C,(\sigma)}(\rho,\phi,z)=\int_{0}^{\theta_m}d\theta\sin\theta \cos\theta_{\rm w} \sqrt{\cos \theta}\,e^{-\gamma^2\sin^2\theta}T(\theta)d_{m,\sigma}^{j}(\theta_{\rm w})g^{(\sigma)}_{m}(\rho,\phi,\theta)
\\
\nonumber
\times
\exp\left\{i[\Phi_{\rm g-w}(\theta)+\Psi_{\rm add}(\theta)+k_{\rm w}\cos\theta_{\rm w} z ]\right\} \label{multipole coefficient}
\end{eqnarray}

\begin{eqnarray}
G_{j,m}^{\pm,(\sigma)}(\rho,\phi,z)=\int_{0}^{\theta_m}d\theta\sin\theta\sin\theta_{\rm w}
\sqrt{\cos \theta}\,e^{-\gamma^2\sin^2\theta}T(\theta)d_{m\pm1,\sigma}^{j}(\theta_{\rm w})g^{(\sigma)}_{m}(\rho,\phi,\theta)\\
\nonumber
\times \exp\left\{i[\Phi_{\rm g-w}(\theta)+\Psi_{\rm add}(\theta)+k_{\rm w}\cos\theta_{\rm w} z ]\right\}.
\end{eqnarray}
\end{widetext}

\section{A short guide to absolute calibration}

An important application of  absolute calibration is the possibility of designing the optical trap  to meet some specific requirement. 
The  parameters required for the determination of the trap stiffness  \cite{SM2}
include the microsphere radius $a$ and refractive index, the laser wavelength $\lambda$ (in vacuum) and power  at the objective entrance port, 
 the refractive indexes of the glass slide $n$ and  of the liquid filling the sample $n_{\rm w}$ (water in many cases), and the objective numerical aperture $\rm NA$
  and transmittance.  All these parameters are usually readily available, except for the last one, which can be reliably measured by the dual objective method \cite{Viana2006}, or  by using a
mercury microdroplet as a microbolometer \cite{Viana2002}.  

One can enlarge the beam waist $w_0$ so as to increase the trapping stability region by overfilling the objective entrance port. 
In a given setup, $w_0$ 
 can be inferred by measuring the laser power transmitted through a diaphragm as a function of its radius, or 
alternatively by imaging the laser beam spot with a CCD \cite{NathanPRE}. 

Once 
these basic input parameters are known, the path to absolute calibration depends on the ratio $a/\lambda$ as follows:

\begin{itemize}

\item $a<\lambda.$
Astigmatism and  interface spherical aberration should be taken into account. The latter is controlled by starting with the trapped bead 
at the very bottom of the sample. One then displaces the objective by a given amount $d.$ Our code  \cite{SM2}
 calculates the resulting spherical aberration effect. Since we neglect reverberation and the contribution of evanescent wave components, 
reliable results are expected in the range
 $d>3\lambda.$
 
When trapping the  small microspheres typically employed in quantitative applications, it is also essential to characterize the 
 astigmatism axis orientation and amplitude. 
For instance, for $a/\lambda\sim 0.25,$ Fig.~4 shows that a small amount of astigmatism leads to a significant 
reduction of the transverse stiffness.

By
imaging the reflected laser spot in a CCD, it is straightforward to measure the axis orientation. The amplitude can be derived by 
fitting the variation of the intensity at the spot  center  with the 
position of the mirror (see Sec.~III for details).

\item $\lambda<a<2\lambda.$
For bead radii $a>\lambda,$ the effect of astigmatism on the trap stiffness is small (see Fig.~9). 
Thus, depending on the required accuracy, 
the stiffness can be calculated using our code as if the trapping beam were stigmatic.  Moreover, the dependence on $d$ is  also negligible
provided that the bead is trapped far from the glass surface. 

\item  $a>2\lambda.$
Our code   is not optimized for very large radii, so we do not recommend its use in this case. 
On the other hand, geometrical optics provides an excellent approximation to the transverse stiffness in this range.
 In this regime, the stiffness is virtually independent 
of wavelength, polarization and trapping height (again as long as reverberation is negligible): $k_{x,y}/P=C/a,$ with the coefficient $C$ independent of $a.$
 For overfilled  oil-immersion high-NA objectives, we find \cite{Dutra} $C=1.1\,{\rm pN}/(\mu{\rm m}\cdot {\rm mW})$ (with $a$ measured in $\mu{\rm m}$)
in the most common case of polystyrene beads in water. 

\end{itemize}


\begin{thebibliography} {99}
\bibitem{footnote1} The displacement theorem \cite{Born&Wolf} is exact only in the paraxial limit. Within the non-paraxial 
formalism developed here, curvature of field leads to a very small modification of stiffness, and thereby to a  small violation of the symmetry property 
discussed here.
\end{thebibliography}
\end{document}